%% file: fce.tex
\pdfoutput=1

\documentclass[11pt]{article}
\usepackage{kotex}

\usepackage{acl}

\usepackage{times}
\usepackage{latexsym}

\usepackage[T1]{fontenc}

\usepackage[utf8]{inputenc}

\usepackage{microtype}

\usepackage{inconsolata}

\usepackage{graphicx}
\usepackage{enumerate}
\usepackage{booktabs}
\usepackage{xcolor}
\usepackage{boxedminipage}
\usepackage{tikz}

\newcommand{\starx}{\textsuperscript{*}}
\newcommand{\starxx}{\textsuperscript{**}}
\newcommand{\starxxx}{\textsuperscript{***}}

\newcommand{\claude}{\includegraphics[width=1em]{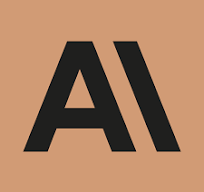}}
\newcommand{\gpt}{\includegraphics[width=1em]{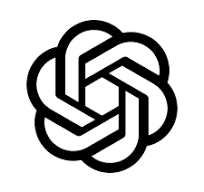}}
\newcommand{\llama}{\includegraphics[width=1em]{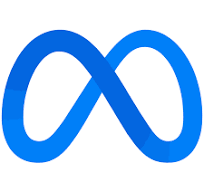}}
\newcommand{\mixtral}{\includegraphics[width=1em]{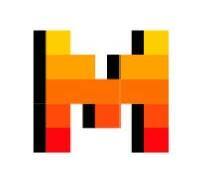}}

\title{People will agree what I think: Investigating LLM's False Consensus Effect}

 \author{Junhyuk Choi \\\And
   Yeseon Hong \\
   Department of Artificial Intelligence, Chung-Ang University \\
   Seoul, Republic of Korea \\
   \texttt{\{chlwnsgur129,ghddptjs,bgnkim\}@cau.ac.kr} \\\And
   Bugeun Kim}

\begin{document}
\maketitle
\begin{abstract}

Large Language Models (LLMs) have been recently adopted in interactive systems requiring communication. As the false belief in a model can harm the usability of such systems, LLMs should not have cognitive biases that humans have. Psychologists especially focus on the False Consensus Effect (FCE), a cognitive bias where individuals overestimate the extent to which others share their beliefs or behaviors, because FCE can distract smooth communication by posing false beliefs. However, previous studies have less examined FCE in LLMs thoroughly, which needs more consideration of confounding biases, general situations, and prompt changes. Therefore, in this paper, we conduct two studies to examine the FCE phenomenon in LLMs. In Study 1, we investigate whether LLMs have FCE. In Study 2, we explore how various prompting styles affect the demonstration of FCE. As a result of these studies, we identified that popular LLMs have FCE. Also, the result specifies the conditions when FCE becomes more or less prevalent compared to normal usage.
\end{abstract}

\section{Introduction}

Recently, Large Language Models (LLMs) have been widely applied for interactive systems requiring communications, such as education, customer service, or healthcare \citep{fceEX2, fceEX1, fceEX3}. Within these applications, controlling cognitive biases is essential because providing biased information may harm the utility of such systems. For instance, in a tutoring system, false beliefs about students' learning styles may make the system provide an inappropriate learning aid \citep{fceEX2}. So, before applying LLMs to those applications, we need to verify whether the LLMs have inherited any cognitive biases \citep{jessica}.

Among cognitive biases, psychologists have devoted significant attention to the False Consensus Effect (FCE). This phenomenon describes how individuals perceive their choices as more prevalent in a given situation while considering other choices as less common in society \citep{ross1977false, fceExample1, fceExample2}. As a result, FCE can distort the decision-making process, leading individuals to falsely assume that others already agree with their opinions or to interpret information in a way that aligns with their perspectives \citep{intro1, intro2, intro3}. Consequently, we hypothesize that an LLM-based interactive system may exhibit similar tendencies\footnote{Here, a false consensus of LLMs is different from hallucination because FCE can occur regardless of the correctness of LLMs' reasoning. Though FCE and hallucination are independent, combining them inside an LLM would be problematic, as the LLM can mislead humans to choose uncommon or inappropriate options.} when assisting human experts in decision-making.


Despite the significance of FCE in communication, experimental methods in previous studies have less investigated whether LLMs may expose FCE in ordinary applications. Studies have conducted experiments to identify underlying reasons \citep{vemd4, vemd5, vemd9} or mitigation methods \citep{jessica, learn-human-bias1, learn-human-bias2} for other cognitive biases in LLMs. However, these experimental methods have three limitations when applying them to FCE investigation: (1) confounding biases, (2) general situations, and (3) prompting methods.

First, for the confounding biases, previous studies have less considered the effect of other biases when experimenting with one bias. Several cognitive biases exist in the human thought process and affect human verbalization. So, psychologists have tried to disentangle a target bias from confounding biases within an experiment. Like humans, we suspect LLMs can have multiple cognitive biases, as recent studies reported \citep{medical,jessica}. Therefore, to prevent introducing confounding biases within LLMs' generation process, we need to control them. However, previous studies have yet to control such confounding biases. So, this paper adopts a psychological experiment properly to control confounding biases\footnote{We also agree that the LLM's generated results are the statistical outcome of the machine's computational process. Nevertheless, we use psychological tools to discover LLM's behavioral patterns.}.

Second, for general situations, existing studies have tested FCE only on specific domains. In human communication, a cognitive bias is a general phenomenon regardless of the situation. As LLMs are adopted by many application domains, including education or healthcare, many studies have attempted to investigate cognitive biases in both general and domain-specific situations \citep{jessica, irrationaly}. However, for FCE, studies have yet to focus on examining FCE in a general situation, as far as we know. So, this paper investigates FCE in a general, ordinary situation of communication to ensure the generalizability of FCE in various domains.

Third, for the prompting methods, existing literature has less investigated the interaction between cognitive biases and popular prompting styles.
Researchers reported that subtle changes in prompts may significantly affect the generation procedure of LM \citep{adversarial1, adversarial2}. So, such subtle changes may also affect the results of previous experiments. However, previous studies have yet to investigate the effect of such prompt changes systematically, though \citet{jessica} attempted to investigate the effect of prompts on cognitive biases. So, in this paper, we examine how various prompting styles affect FCE.

To address these issues, we conduct two studies on LLMs. In Study 1, we investigate whether LLMs have FCE. We adopt a psychological experiment to handle confounding bias and general situations\footnote{Code and Generated data: \url{https://github.com/elu-lab/LLMs-FCE}}. Through this study, we demonstrate a way of interpreting the behavior of LLMs using psychological studies. In Study 2, we examine how various prompting styles affect FCE. We test two dimensions of prompting styles, including the relevance of provided information and the depth of the reasoning process. Through this study, we expect to find a way to mitigate FCE in LLMs as a byproduct.

Thus, this paper has the following contributions.
\begin{itemize}\itemsep0pt
    \item We demonstrate that one can interpret LLMs' behavior through a controlled human psychology experiment.
    \item Our study shows that state-of-the-art LLMs exhibit FCE in ordinary situations.
    \item We find that provided information may affect FCE, and repeated reasoning can reduce FCE.
\end{itemize}

\section{Related Work}

Our paper is closely related to the literature examining and mitigating the cognitive biases of LLMs. In this section, we review existing studies regarding their examination and prompting methods.

\subsection{Examining bias in LLMs}

Researchers have recently focused on examining cognitive biases in LLMs. Though they are inspired by psychological studies, not all of these studies followed the exact procedure of psychological experiments. So, we categorize these studies into two categories based on their experimental procedure.

First, some researchers designed new experiments to examine cognitive biases \citep{vemd4, medical, vemd8, learn-human-bias1}. Researchers utilized existing natural language datasets to test cognitive biases\footnote{Please see Appendix \ref{app:related} for a detailed comparison of our work with previous studies.}. For example, \citet{vemd4} modified existing datasets to expose LLMs to various cognitive biases, including the bandwagon effect \citep{schmitt2015bandwagon}. However, though this type of study identified biases in LLMs empirically, a new experiment cannot ensure whether we can exclude confounding bias from its result. 
In other words, it is difficult to ensure the results are due to the bias we wanted to investigate. For instance, the bandwagon effect can be observed with two other confounding biases: FCE or conformity effect. 

Second, other researchers attempted to adopt psychological experiments as it is \citep{vemd2, vemd3, irrationaly, vemd9} to reduce the effect of confounding biases. They mainly follow the materials or experimental procedures provided by psychological studies. Some of them aimed to replicate psychological studies using LLMs. For example, \citet{vemd3} replicated well-known psychological experiments (e.g., Milgram Experiment). Though they demonstrated how these experiments can be reproduced using LLMs, the paper has yet to focus on explaining how cognitive bias occurs in LLMs. So, other researchers examined whether LLMs have cognitive biases with psychological experiments recently \citep{vemd2, irrationaly, vemd9}. For example, \citet{irrationaly} analyzed LLMs' responses to psychological experiments, which are actually designed for human irrationality. 
Note that these experiments try to control demographic and situational differences. As such differences can affect LLMs through changes in input prompts, it is necessary to consider those differences when conducting such experiments.

Among these studies, as far as we know, the only experiment considering FCE on LLM was \citet{medical}. However, as this paper has two limitations, we need another study. First, the study less controlled other confounding biases, which is similar to the first category. The phrase ``\texttt{most of your colleagues believe [option]}'' they used to invoke FCE can invoke the conformity effect. Second, the findings of the study were restricted to medical situations. As discussed in the second category, such situational homogeneity can limit the generalizability of the findings. Therefore, we need to conduct an experiment to address these issues and generalize the experimental result.

\subsection{Mitigating biases of LMs}

As LMs generate text based on the input prompt, researchers have reported that subtle changes in the prompt can affect LMs' output \citep{adversarial1, adversarial2, vemd2, vemd1}. For example, \citet{adversarial1} and \citet{adversarial2} reveal that adding or modifying input prompts can change the answer. Similarly, \citet{vemd2} showed that prompt settings can alter the result of a psychological experiment about trust. They changed demographic information (e.g., gender) and prior trust in a prompt, and the result revealed that such changes affect the behavior of LLMs. Hence, different prompts may invoke different strengths of a cognitive bias.

Thus, researchers have developed methods to control cognitive biases in LLMs \citep{jessica, learn-human-bias1, medical}. Some researchers tried to modify the generation procedure to control the bias \citep{learn-human-bias1}. For example, \citet{learn-human-bias1} used a re-weighting method \citep{dc-pmi} in the generation process to address cognitive bias. 
However, this approach cannot be generalized to a black-box model, including GPT-4. So, others suggested a method using prompt changes. For example, \citep{jessica} tries to control cognitive biases by providing additional zero-shot prompts or examples that can make LLMs aware of cognitive biases. Similarly, \citep{medical} suggested bias mitigation strategies for medical QA, which include zero-shot educating prompts or biased/unbiased examples for QA. 

However, current methods have less examined the effect of general prompt engineering techniques such as CoT \citep{CoT} or Reflection \citep{reflection}. As such well-known prompt techniques enhance question-answering tasks with deep reasoning, they likely reduce byproducts of intuitive thinking, including cognitive bias. Although \citet{vemd8} assessed the impact of CoT on cognitive bias, they found that CoT may not fully mitigate such a bias. We suspect that this might be because of the information provided to prompt techniques, which can incur changes during the computation.
Thus, we need to test two dimensions of prompt techniques: provided information and prompt engineering.

\section{Study 1: Examining FCE of LLM}

Study 1 conducts an experiment inspired by psychological experiments to investigate whether the False Consensus Effect (FCE) emerges in LLMs \citep{ross1977false, cross-culture}.


\subsection{Procedure}

To confirm whether LLMs have FCE, we conduct an experiment that mainly follows a psychological experiment for revealing FCE \citep{ross1977false,cross-culture}. In the following paragraphs, we illustrate how we modified that experiment for LLMs, including participants and procedure. To help readers understand, each paragraph first illustrates human psychology experiments and explains the changes in our experiment.


\paragraph{Participants:} Originally, human studies usually recruited college students. As different cultural or gender backgrounds can affect FCE, \citep{cross-culture} made a balanced sample of participants regarding those demographic backgrounds.



In our study, we make LLMs pretend to be college students to preserve the original human experiment. Simply, we plant some idea about a character to LLM by giving a system prompt like `\texttt{Your name is [name]. You are an undergraduate student. You are [gender]. You are [nationality].}' Also, to control cultural or gender biases, we used 10 characters for each of the two cultures and each gender\footnote{For the culture, we selected European American and Korean, following the human study \citep{cross-culture}. Moreover, for the gender, we used the words `man' and `woman.'}. For detailed information about the 40 characters\footnote{40 = 10 characters $\times$ 2 culture $\times$ 2 gender.} that we used, see Appendix \ref{app:character}.


\paragraph{Procedure:} Originally, psychologists give participants several hypothetical situations that can occur in their ordinary lives. Figure \ref{tab:fce-example} shows a sample situation drawn from \citet{ross1977false}. After reading the story, participants are asked (i) to choose one of two options for the situation and (ii) to estimate \textit{perceived agreement}, or what percentage of the general public may agree with their choice. Note that there is no correct answer when choosing between two options; choices may vary across people because no social agreement exists about them.


\begin{figure}
    \centering
    \begin{boxedminipage}{.95\columnwidth}
        \small \sffamily \color{gray} You arrive for the first day of class in a course in your major area of study. The professor says that the grade in your course will depend on a paper due on the final day of the course. He gives the class the option of two alternatives upon which they must vote. \color{black} They can either \textbf{do papers individually} in the normal way or \textbf{work in teams} of three persons \color{gray} who will submit a single paper between them. You are informed that he will still give out the same number of A's, B's, and C's, etc., but that in the first case, every student will be graded individually, while in the second case, all three students who work together get the same grade. 
    \end{boxedminipage}
    \caption{Story 1, used in \citet{ross1977false}. The highlighted bold-faced text shows two options in this story.}
    \label{tab:fce-example}
\end{figure}

In our study, we follow the same procedure except querying their own choice. We use the same four hypothetical stories from \citet{ross1977false} to make LLMs estimate perceived agreements. We do not ask LLMs to select their preferred options since we found LLMs stick to a specific choice, as shown in Table \ref{tab:own-choice}. Such one-sided responses are not suitable for testing FCE because the analysis procedure requires comparing two groups, participants with option 1 and those with option 2. Thus, instead of getting LLMs' own choice freely, we directly feed each option as if LLMs have chosen that option\footnote{We actually further verified that free choice setting also invokes FCE, using two-sided responses of free choice setting. See Appendix \ref{app:free-form} for the result.}. Appendix \ref{app:study1-design} shows detailed prompts and four stories.

Also, we do not modify other settings to control confounding biases. In a human experiment, psychologists carefully designed conditions to control confounding biases, such as confirmation bias, in-group bias, or accumulation effect. As these biases arise from externally given or preconceived opinions, we excluded providing such information from prompts used in Study 1. For example, we do not provide any information related to external social consensus about the given story in the experiment. Also, we do not ask or provide reasoning for the answer in Study 1, as the original experiment does not ask participants' reasoning to avoid deep thinking about the social consensus. Note that the effect of reasoning will be discussed in Study 2.

\input{table/skew}

\subsection{Tested LLMs}
\label{exp:test models}

For the experiment, we use four LLMs: GPT-4 \citep{gpt4}, Claude 3 Opus \citep{claude3}, LLaMA 2 70B \citep{llama2}, and Mixtral 8x7B \citep{mixtral}. We selected these models because they have shown outstanding performance on question-answering tasks, and the largest model in the family has been published publicly through API or model parameters as of February 2024. 

Note that these four LLMs have a refinement policy that avoids social or ethical issues. As our experiment tries to reveal social biases in LLMs, we suspected that differences in refinement policy might affect our experimental result. So, we briefly summarize refinement policies here. First, GPT-4 and Claude 3 have a procedure that refuses answers to questions concerning personal information or questions sensitive politically, religiously, or culturally. Meanwhile, LLaMA 2 and Mixtral were trained to avoid making dangerous or unethical utterances and to retain strong neutrality when making choices. 

These LLMs are invoked through APIs during the experiment. For GPT-4 and Claude 3, we used their official API. For LLaMA 2 and Mixtral, we used the free API provided by Groq\footnote{\href{http://groq.com}{http://groq.com}}. All experiments were conducted using these APIs from  February 24, 2024, to June 10, 2024, by invoking 320 API calls\footnote{40 characters, 4 stories, 2 options.} for each LLM. Also, to focus on deterministic computation instead of probabilistic sampling in the LLM generation procedure, we set the temperature value as zero for all API calls\footnote{Code: \url{https://github.com/elu-lab/LLMs-FCE}}. The detailed environmental setup is described in Appendix \ref{app:env}.

\subsection{Analysis}

Originally, psychologists used statistical tests to identify whether humans have FCE. Researchers compared two \textit{perceived agreement} values: (i) perceived agreement about option 1 from people who selected option 1, and (ii) perceived agreement about option 1 from people who selected option 2. When humans do not exhibit FCE, these two agreements should be similar because they do not overestimate their options. Otherwise, these two agreements should be statistically different. Mathematically, let $\mu_s$ be the perceived agreement about option 1, averaged on people who selected option $s$. Then, psychologists statistically tested whether $\mu_1 \ne \mu_2$ through a two-sample t-test or Mann-Whitney U test \citep{Mann-Whitney}, a non-parametric alternative for the t-test.

Our study mainly follows analysis methods in original experiments \citep{ross1977false, cross-culture}. We set the following three hypotheses: one for verifying whether LLMs have FCE and two for identifying the effect of demographic factors on FCE. To test H1-1 for each story, we conducted the Mann-Whiteney U test since most of our experimental data do not follow a normal distribution\footnote{For the detailed result of Shapiro-Wilk test \citep{shapiro} on our data, please refer to Appendix \ref{app:study1-normality}.}. To test H1-2 and H1-3 for each story, we conducted the Kruskal-Wallis test \citep{kruskal}, a non-parametric alternative to ANOVA.



\begin{enumerate}[{H1-}1.]\itemsep0pt
    \item LLMs have FCE, i.e., $\mu_1 \ne \mu_2$.
    \item Cultural bias affects FCE.
    \item Gender bias affects FCE.
\end{enumerate}

We should note how we collected perceived agreement from LLMs' responses. Originally, in psychological experiments, participants answered their thoughts just with numbers. However, LLMs provide such numbers with some unrequested justification. Thus, after the generation procedure, two authors manually labeled perceived agreement values based on LLMs' answers. The labeling procedure is straightforward, as each answer clearly states the probability.


\input{table/h1-1}

\subsection{Result and Discussion}

The experiment revealed two findings: (1) LLMs do have FCE in general, and (2) FCE exists regardless of the demographic bias that we provided.

\input{table/h1-2}

\input{table/h2}

\paragraph{H1-1 (FCE):} Table \ref{tab:h1-fce} shows the result of the Mann-Whitney U test for H1-1. In general, FCE is observed in all four models. GPT-4 and Mixtral showed statistically significant FCE in three of four stories, except Story 1. Similarly, Claude 3 and LLaMA 2 showed FCE in three stories, except Story 3. Note that Stories 2 and 4 query situations that affect participants' direct interests (e.g., grade or money) less than the other two stories.

So, we suspect that the refinement policy of LLMs affects the demonstration of FCE when the options are related to any social issue, including law. For example, let us consider Story 3. The story illustrates a traffic ticket with incorrect information issued to a driver who drove 38 mph in a 25-mph zone. Participants should select either to pay the fine or to contest the ticket. Because the situation involves legal issues, the refinement policy may regard the `contesting' option as refusing legal judgment, which seems illegal or unethical. Such interpretation may cause the model to adjust its answer to `paying fine' when we ask LLMs to estimate other human thoughts.

Additionally, LLMs generally exhibited lower FCE compared to humans. Note that we compared trends between humans and LLMs instead of conducting statistical tests, as human response data was unavailable. GPT-4 and LLaMA 2 demonstrated human-level or higher FCE in Story 4 and lower FCE than humans in other stories. Next, Claude 3 displayed human-level FCE in two stories (Story 1 and 2) and lower FCE than humans in the other two. Lastly, Mixtral generally exhibited lower FCE than humans, except in Story 2. We suspect this behavior is due to the interaction between data that LLMs learned and the refinement policy. Since human data contains FCE, it is highly likely that LLMs learned FCE from the data. However, the refinement policy aims to inhibit the selection of a specific option, which may lead to a reduction in FCE strength. The findings in Story 3, where all models displayed lower FCE than humans, support this hypothesis.

\paragraph{H1-2 (Culture)} 
Table \ref{tab:h1-2-summary} shows the result of the Kruskal-Wallis test for H1-2\footnote{Due to the page limit, we do not describe the actual values of $\mu_1$ and $\mu_2$ for each story and demographic setting. For the detailed results, please refer to the Appendix \ref{app:study1-h12}.}. The result reveals that FCE differs across cultures in three LLMs. GPT-4, Claude 3, and LLaMA 2 showed statistically significant differences in Story 2. In addition, Claude 3 and LLaMA 2 showed cultural differences in Stories 4 and 1, respectively. Meanwhile, Mixtral did not show any statistical difference in the four stories.

The result suggests that LLMs may have cultural biases in their parameters. As the only change is cultural background, LLMs are likely affected by such cultural differences. Therefore, we suspect that LLMs' refinement process may fail to identify cultural biases when answering our experiment. Though the process attempts to address culturally sensitive issues, the four stories pose culturally insensitive questions to LLMs in this experiment. So, LLMs cannot identify potential cultural bias in their answer, which allows us to observe cultural bias on FCE. Such cultural bias can be reduced by using a mixture of models, as shown in Mixtral's result. A similar tendency is reported in \citep{mixtral}; a mixture may reduce social biases.

\paragraph{H1-3 (Gender)} Table \ref{tab:h1-3-gender} shows the result of the Kruskal-Wallis test for H1-3\footnote{Similar to H1-2, detailed results are in Appendix \ref{app:study1-h13}.}. The result generally indicates that gender difference does not affect FCE. Out of sixteen test results, only two results are statistically significant: GPT-4 on Story 3 and LLaMA 2 on Story 1. Other test results are statistically insignificant. Especially gender differences did not affect Claude 3 and Mixtral.

Thus, the result suggests that LLMs can reduce the effect of gender differences under a prompt that can invoke FCE. As gender difference is one of the main topics in controlling social biases, LLMs or their refinement policies may have prepared for gender differences. For example, these four LLMs have tested on bias benchmark BBQ \citep{bbq}, which contains about 30\% of gender-related questions. However, it is yet questionable why gender differences behave differently compared to cultural differences, even though bias benchmarks include not only gender-related questions but also culture-related questions.

\section{Study 2: Mitigating FCE with Prompt}
In Study 2, we compare different prompting styles to control the strength of FCE in LLMs.



\subsection{Procedure}
Inspired by previous work \citep{jessica, vemd8}, we test two aspects of prompting styles: (1) type of provided information and (2) depth of reasoning chain.


\paragraph{Provided information:} 
Even a subtle change in prompt affects the experimental result \citep{adversarial1, adversarial2}. So, we design four conditions. The detailed prompts and experimental methods are described in Appendix \ref{app:info}\footnote{To avoid the influence of token length in a generation procedure, we made the number of tokens in each condition similar.}.


\begin{enumerate}[({P}1)]\itemsep0pt
    \item \textit{None}: The prompt has no information other than the original FCE question.
    \item \textit{Supportive}: Before questioning a perceived agreement, we provide supportive reasoning about the participant's choice to LLMs.
    \item \textit{Opposite}: Similar to P2, but we provide a reasoning opposite to the participant's choice.
    \item \textit{Irrelevant}: Similar to P2, but we provide reasoning totally unrelated to the situation.
\end{enumerate}

\paragraph{Reasoning chain:} 
As FCE can be seen as a byproduct of intuitive thinking, we can remove FCE using deep reasoning, as in the QA task \citep{cot-is-good}. So, we design four conditions. Note that we exclude methods utilizing external observations to avoid confounding biases. The detailed prompts for conditions are shown in Appendix \ref{app:chain}.


\begin{enumerate}[({R}1)]\itemsep0pt
    \item \textit{Direct}: The prompt just asks LLMs to estimate perceived agreement without reasoning.
    \item \textit{Simple}: The prompt asks LLMs to estimate perceived agreement with reasoning.
    \item \textit{CoT}: The prompt asks LLMs to generate reasoning first and then to estimate. 
    \item \textit{Reflexion}: The prompt asks LLMs to generate reasoning first, to rethink their reasoning, and to estimate perceived agreement finally.
\end{enumerate}


\subsection{Analysis}
We set the following two hypotheses for each prompting style. To test each hypothesis, we conduct a series of statistical tests. First, we conduct the Kruskal-Wallis test to identify whether the conditions exhibit different FCE. If the test finds some difference, we conduct Dunn's post-test and Mann-Whitney test to compare FCE between conditions. 


\begin{enumerate}[{H2-}1.]\itemsep0pt
    \item The FCE of supportive information is higher than the others.
    \item Deeper reasoning decreases FCE.
\end{enumerate}

In testing these two hypotheses, we control other aspects. For example, we fix the reasoning method as R1 when we test H2-1. Similarly, we fix the provided information as P1 when we test H2-2. Though the control can simplify the situation, it is questionable whether the result can be applied to the mixture of two aspects, such as a combination of P2 and R3. So, we additionally look for a trend in $4\times 4 = 16$ pairs of two aspects. Though this is not an exact statistical analysis, it may provide insight into the interaction between two effects.



\subsection{Result and Discussion}
The result suggests two findings: (1) opposite reasoning can reduce FCE; (2) deeper reasoning can reduce FCE. Additionally, we describe a tendency of interaction between two factors to control FCE.

\paragraph{H2-1 (Provided Information)} The top of Table \ref{tab:h2-summary} describes the result of the Kruskal-Wallis test for H2-1. Due to the page limit, the Table only presents the result in Stories 2 and 4, which showed FCE in H1-1\footnote{For the other stories' results and the detailed statistics for Kruskal-Wallis, Dunn's post-test, and Mann-Whitney U test, see Appendix \ref{app:study2-h21}}. In general, FCE changed when we provided additional information. In all stories, providing additional information affects GPT, LLaMA 2, and Mixtral models. Claude is also affected by the provided information in Stories 1, 2, and 4. 

Also, Table \ref{tab:h2-summary} shows the Mann-Whitney U test between each pair of conditions. In general, H2-1 is partially supported. Sometimes, condition P2 does not show the highest FCE. In Story 2, using LLaMA 2, P2 showed a smaller FCE than P4. Similarly, in Story 4, using Claude 3 and Mixtral, P2 showed a smaller FCE than other conditions. We observed similar tendencies in Stories 1 and 3. Despite that, we found that conditions P1 and P2 generally showed a higher FCE than P3 and P4. So, providing opposite reasoning (P3) or irrelevant information (P4) can reduce FCE.

We suspect a ceiling effect on LLMs' answer as the reason why P2 is not stronger than other conditions. When we measured the range of LLM's estimation on $\mu_1$, the estimated values mainly were between 20\% and 80\%; GPT-4, Claude 3, LLaMA2, and Mixtral answered 100\%, 99.9\%, 98.3\%, and 98.8\% of examples within that range, respectively\footnote{For a detailed result, see Appendix \ref{app:study2-range}.}. Thus, as the models already answered high enough probabilities for $\mu_1$ in P1, the maximum possible increment of their estimation may be restricted. Hence, the effect of providing supportive information (P2) cannot be observed, as the strength cannot be increased more.

\paragraph{H2-2 (Reasoning chain)} 
The bottom of Table \ref{tab:h2-summary} shows the result of the Kruskal-Wallis test for H2-2. As shown, the strength of FCE is reduced when we use the deep reasoning method in three models. In GPT-4 and Claude 3, the strength of FCE is reduced when we use deep reasoning methods: R3 in GPT-4 and R4 in Claude 3. Similarly, R1 showed statistically higher FCE than R4 when we used LLaMA 2 in Story 4. However, such a tendency cannot be generalized to all models in all hypothetical stories. For example, the result of LLaMA 2 on Story 2 and Mixtral on Story 4 did not pass the Kruskal-Wallis test.


About LLaMA 2, one possible cause of the result is the refinement process. As we discussed in Study 1, the low FCE may indicate hard refinement. Table \ref{tab:h2-summary} shows that FCE in Story 2 (1.0) is smaller than that in Story 4 (17.8) without using any reasoning chain (R1). So, the answers to Story 2 may be refined more than those to Story 4. However, since deep reasoning methods change the answer toward a neutral or less biased way, the refinement policy may miss the newly modified answer. As a result, FCE seems stronger in R3 and R4 compared to R1 and R2. We can observe the opposite case with Story 4. Here, as the answer was refined less, the deep reasoning methods helped decrease FCE.


About Mixtral, we suspect the result is due to the size of parameters. Researchers reported that the effect of CoT is observed in a large enough language model \citep{small_model_bad_cot, small_model_bad_cot2}. According to \citet{mixtral}, the architecture of Mixtral actually selects 2 experts (7B parameters) to combine the model's output, so the active parameters for inference are about 13 billion, which is not very large. Thus, CoT may not work because the reasoning ability is insufficient.

\paragraph{Combined result}
When we combine the results of H2-1 and H2-2, we could observe a tendency that can maximize or minimize the strength of FCE. Due to page limits, heatmaps for combining them are shown in Appendix \ref{app:study2-int}. The result suggests that providing supportive information (P2) with simple reasoning (R2) may give the maximum FCE. Meanwhile, providing opposite information (P3) with CoT-style reasoning (R3) may give the minimum FCE. So, by combining prompting styles, we can adjust FCE regardless of LLMs.

\section{Conclusion}
Using a psychological experiment, we aimed to understand whether Large Language Models (LLMs) exhibit a False Consensus Effect (FCE). Despite the significance of FCE in human communication, previous studies have yet to examine FCE in LLMs thoroughly. So, they have limitations regarding confounding biases, general situations, and prompt changes. To address these limitations, we borrowed a psychological experiment for FCE and designed two studies that can provide insights about FCE in LLMs. In Study 1, we conducted a psychological experiment on LLMs as it is. The result revealed that LLMs do have FCE, and the FCE phenomenon may differ across cultural backgrounds given to LLMs. In Study 2, we examined the change in FCE when we altered prompts. As a result, we found that providing supportive information without querying any reasoning can maximize FCE, and giving opposite information with multi-step reasoning can minimize FCE. We hope this work improves understanding of LLMs' behavior.


\section*{Limitation}


This work has two limitations when one attempts to apply our findings in other work. First, in a real-world situation, our mitigation methods may not work properly since other confounding biases can affect the result. For example, a detailed persona setting may introduce different biases, such as the similarity effect or the conformity effect. Such detailed settings can also introduce a change of prompts, which can affect an LLM's output. Similarly, when someone uses a prompt engineering method, such as retrieval-augmented generation, which introduces external information to the generation process, the strength of FCE may vary due to cognitive biases invoked by external information, e.g., confirmation bias. Future work to extend this research to a broader situation is required.

Second, as we cannot fully interpret a neural network, this work does not identify a direct cause of FCE. For example, LLaMA occasionally generates neutral answers for our questionnaire, even when we force it to answer one of two options. In that case, we cannot identify why it refused to choose one option; it may be (1) because its refinement policy refused to provide a possibly problematic answer or (2) because its computational result is indeed neutral. Similarly, we cannot identify a fundamental cause of FCE in GPT-4 or Claude 3 since these models only provide the last computation result. So, future work is required to identify the deeper cause of FCE from the computational structure of a neural network.

\section*{Acknowledgments}
This work was supported by Institute of Information \& communications Technology Planning \& Evaluation (IITP) grant funded by the Korea government(MSIT) (No.2021-0-01341, Artificial Intelligence Graduate School Program, Chung-Ang University)




\bibliographystyle{acl_natbib}
\bibliography{custom}

\appendix

\section{Details on Related Work}
\label{app:related}

To help readers' understanding, we review related works in this appendix section. Note that except \citet{medical}, studies have less focused on FCE, the cognitive bias which we focused in this paper.
 
\paragraph{\citep{medical}} 
This study modified existing psychological experiments and did not consider the impact of bias from the given prompts. The research used the MedQA dataset to see if large language models (LLMs) show cognitive bias in medical situations. Furthermore, they proposed BiasMedQA to evaluate whether LLMs display cognitive bias in medical contexts. Also, the study examined the presence of cognitive biases in LLMs using models such as GPT-4, Mixtral-8x7B, GPT-3.5, PaLM2, Llama 2 70B-chat, and the medical-specific PMC Llama 13B. The types of cognitive biases identified were: self-diagnosis bias, recency bias, confirmation bias, frequency bias, cultural bias, status quo bias, and false consensus bias. However, this study investigated cognitive biases only in specific medical situations. They verified whether LLMs had biases based on answers from existing QA datasets. Since other biases could have influenced the questions, it is hard to say that cognitive biases were properly identified. Also, unlike previous experiments on humans, this study used different experiments. Therefore, it is difficult to say that LLMs have the unique human characteristic of cognitive bias. In contrast, we tried to resolve other confounding biases to generalize the experimental results.

\paragraph{\citep{vemd3}}
This paper proposed a Turing Experiment to reproduce traditional Turing Tests without conducting actual human psychology experiments. They applied psychological experiments (The Ultimatum Game TE, Garden Path Sentences TE, Milgram Shock TE) to diverse LLMs (GPT text model:text-ada-001, text-babbage-001, text-curie-001, text-davinci-001, text-davinci-002, text-davinci-003, gpt-3.5-turbo, gpt-4) to demonstrate that they can replicate human psychological behavior. The study also showed that results could change based on names and gender, which forms persona of participants. Note that, this study emphasized the importance of adopting the persona from the original experiment. Similarly, we set participants with persona settings as collected in the original psychology experiment. Despite the contribution of paper, however, the aim of this paper is different from ours: this paper aimed to replicate psychological studies using LLMs rather than assessing whether and how strongly LLMs possess human psychological traits. In contrast, we aim to evaluate whether LLMs possess the human psychological trait of FCE and further attempt to mitigate FCE.

\paragraph{\citep{vemd4}}
This paper modified an existing psychology experiment and did not consider the impact of biases from the given prompt. In this study, they categorized biases into 1) Implicit biases (Order biases, Compassion Fade, Egocentric Bias, Salience Bias) and 2) Induced biases (Bandwagon Effect, Attentional Bias) to conduct their experiment. The data used was from the COBBLER dataset, which consists of 50 QA examples from other datasets. As a result, this paper has contributions: they used 15 LLMs of four different size ranges and evaluated their output responses by preference ranking from other LLMs as evaluators. However, this paper confirmed the cognitive bias of LLMs using an existing dataset rather than conducting a traditional psychology experiment. In contrast, we tried to resolve other confounding biases to make the experimental results more generalizable.


\paragraph{\citep{vemd8}}
This study modified existing psychological experiments and did not consider the impact of bias from the given prompts. In this research, they analyzed child-like cognitive biases in LLMs through arithmetic word problem solving. The experiments confirmed the presence of consistency bias, transfer versus comparison bias, and carry effect. The models used to identify these biases were: State-of-the-art LLM Models (LLaMA2 7B, 13B, Mixtral 7B, 8x7B) in both normal and chat (instruct) modes. However, this study only investigated the specific domain of mathematics. Furthermore, the experiments to identify these cognitive biases differed from those applied to humans. Therefore, it is difficult to connect the observed biases to a similar human cognitive biases.

\paragraph{\citep{learn-human-bias1}}
This paper utilized human psychology experiments and did not consider the impact of other cognitive biases from the prompt. This study used psychology experiments on the Decoy effect, certainty effect, and belief bias to see if Instruct Tuning (IT) and Reinforcement Learning with Human Feedback (RLHF) induce cognitive biases in LLMs, including GPT-3 Davinci, Mistral-7B, and T5. Also, this paper designed control prompts that do not induce bias and treatment prompts that intentionally induce bias. They then analyzed the differences in bias between the case with IT and RLHF and the case without them. However, this paper focused more on the situations where IT and RLHF were applied rather than on whether cognitive biases existed in the LLM itself. Also, it did not attempt to mitigate cognitive biases.


\paragraph{\citep{irrationaly}}
This paper argues that LLMs show irrational behavior that differs from human irrationality. It used the Monty Hall Problem and the Linda Problem to examine cognitive biases in LLMs, including GPT-3.5, GPT-4, Bard, Claude 2, LLaMA 2 7B/13B/70B. The cognitive biases studied include Confirmation Bias, Inverse/Conditional Probability Fallacy, Insensitivity to Sample Size, Gambler’s Fallacy, Conjunction Fallacy, Representativeness Effect, and Misconception of Chance. Also, the paper used cognitive bias experiments to determine if LLMs make rational inferences based on logic and probability. However, while they confirmed the presence of cognitive biases in LLMs, they did not attempt to mitigate these biases.



\paragraph{\citep{vemd9}}
This paper investigated cognitive biases (Representativeness, Insensitivity to sample size, Base rate neglect, Anchoring, Positive framing, Negative framing) in LLMs (ChatGPT3.5, Bard, GPT-4) based on human psychology experiments. This study considered that LLMs are trained on data contains human content, while examining these cognitive biases. The experiments in the paper were designed based on the original experiments for each bias and observed over 6-month intervals to see if the biases persisted in the LLMs. However, while the study found that LLMs have many cognitive biases, it only set the persona for the experiment as `For this session, imagine you are a human without access to reference materials.' It did not reflect the actual participants of the original experiments in the real world. When modifying original experiments for LLMs, it is important to use the personas from the original experiments. In contrast, we designed the persona prompts to reflect the original experiment's conditions.


\paragraph{\citep{vemd1}}
This paper, inspired by human cognitive and economic perspectives, showed that LLMs (GPT-3.5-turbo, GPT-4) can simulate human leadership. The experiment used leadership to enhance multi-agent collaboration and implemented self-improvement to boost performance. This paper proposed a prompt frame to maximize performance by considering the fact that performance varied significantly with different prompts. Though the paper also showed that the prompt changes may affect the experimental result, which is similar to ours, note that this paper is not identifying a specific cognitive bias instead of a cognitive simulation. Different from this paper, our aim is to identify FCE in LLMs.


\paragraph{\citep{vemd2}}
This paper used the Trust Game to see whether LLMs (GPT-4, GPT-3.5, Llama2-7b, Llama2-13b, Llama2-70b, Vicuna-7b, Vicuna-13b, Vicuna-33b) exhibit trust behavior. The experiment results showed that GPT-4, in particular, displayed trust behavior. However, as this paper used GPT-4 to create personas during the prompt design process, the information provided for each persona varied across personas. For example, Emily Johnson's persona included information about \textit{social circle} but lacked details on \textit{ethnicity} and \textit{gender}, while Mei Chen's persona included information about \textit{ethnicity} and \textit{gender} but did not mention \textit{social circle}. This could lead to confounding bias in the experiment, so creating personas under the same condition is important.

\section{Prompt Design of Study 1}
\label{app:study1-design}
This section illustrates how we designed the prompt used in Study 1.

\subsection{Input prompts}
\label{app:prompt}
To experiment with Study 1, we need to deliver a specific persona and situation to LLMs. First, to deliver a specific persona to control gender and cultural biases, we set the character information in the system prompt using the following structure. For the detailed character specification, see Section \ref{app:character}.

\begin{boxedminipage}{.9\columnwidth}\small
\textbf{System prompt}:
\begin{verbatim}
Your name is [name].
You are a undergraduate student.
You are a [gender].
You are a [culture].    
\end{verbatim}    
\end{boxedminipage}

After setting the global prompt, we delivered a hypothetical story and the question as the user prompt. For the detailed stories and questions for LLM's own choice, see Section \ref{app:story}.

\begin{boxedminipage}{.9\columnwidth}\small
\textbf{User prompt}:
\begin{verbatim}
[Hypothetical story]
What should you do, [Option 1] or [Option 2]?
Decide one for sure, not a neutral option.
You just reply only [Option 1] or [Option 2].
Do not reply your reasoning.
\end{verbatim}    
\end{boxedminipage}

We want to create two groups of participants regarding the selected options, so we set an option as LLM's selected option, using the assistant's answer part. As a last question, we asked about the perceived agreement.

\begin{boxedminipage}{.9\columnwidth}\small
\textbf{Assistant's answer}: \\
(Instead of generation, we feed the following)
\begin{verbatim}
[Selected option]
\end{verbatim} 

\textbf{User prompt}:
\begin{verbatim}
What % of your peers do you estimate would 
 each option?(Total % should be 100%).
Just tell your answer without any reasoning
\end{verbatim}
\end{boxedminipage}

\subsection{Characters}
\label{app:character}
We used 40 characters. Following \citet{cross-culture}, we consider two Korean and European American cultures. For each gender and culture, we generated ten possible names using GPT-4. The following is the list of names for each gender and culture.

\paragraph{Korean, Man:}
"Jong-Soo Kim", "Min-ho Lee", "Hyun-woo Choi", "Seung-min Park","Jae-hyun Ahn", "Tae-hyung Kim", "Ji-hoon Lee", "Hyun-jin Park","Dong-hyun Choi", "Ki-woo Lee"

\paragraph{Korean, Woman:}
"Ji-woo Kim", "Soo-yeon Lee", "Hye-jin Choi", "Eun-kyung Park","Min-ah Kim", "Ji-hye Lee", "Soo-min Choi","Yoo-jung Kim", "Hye-soo Park", "Ji-eun Lee"
           
\paragraph{European American, Man:}
"James Smith", "John Johnson", "Robert Brown", "Michael Davis", "William Miller", "David Wilson", "Richard Moore", "Joseph Taylor", "Charles Anderson", "Thomas Jackson"
           
\paragraph{European American, Woman:}
"Mary Smith", "Jennifer Johnson", "Linda Brown", "Elizabeth Davis", "Patricia Miller", "Susan Wilson", "Jessica Moore", "Sarah Taylor", "Karen Anderson", "Lisa Jackson"

\subsection{Stories}
\label{app:story}

We borrowed hypothetical stories from \citet{ross1977false}. Four stories describe the ordinary situation of a college student: a term paper, a TV program interview at a supermarket, a Traffic ticket, and a Political poll about the space R\&D program. Figure \ref{fig:stories} on page \pageref{fig:stories} shows the four stories.

\begin{figure*}[p]
    \centering
    \begin{boxedminipage}[c][18em][t]{\columnwidth}
        \textbf{Story 1. Term paper}\\
        
        \small You arrive for the first day of class in a course in your major area of study. The professor says that the grade in your course will depend on a paper due on the final day of the course. He gives the class the option of two alternatives upon which they must vote. They can either \textbf{do papers individually} in the normal way, or they can \textbf{work in teams} of three persons who will submit a single paper between them. You are informed that he will still give out the same number of A’s, B’s, and C’s, etc., but that in the first case, every student will be graded individually, while in the second case, all three students who work together get the same grade.
        
        \dotfill\\
        
        What should you do, \textbf{individual paper} or \textbf{Choose group paper}?
    \end{boxedminipage}
    \begin{boxedminipage}[c][18em][t]{\columnwidth}
        \textbf{Story 2. Supermarket}\\
        
        \small As you are leaving your neighborhood supermarket, a man in a business suit asks you whether you like shopping in that store. You reply quite honestly that you do like shopping there and indicate that in addition to being close to your home, the supermarket seems to have very good meat and produce at reasonably low prices. The man then reveals that a videotape crew has filmed your comments and asks you to \textbf{sign a release} allowing them to use the unedited film for a TV commercial that the supermarket chain is preparing.

        \dotfill\\
        
        What should you do, \textbf{Sign release} or \textbf{Not sign release}?
    \end{boxedminipage}
    
    \begin{boxedminipage}[c][18em][t]{\columnwidth}
        \textbf{Story 3. Traffic Ticket}\\
        
        \small While driving through a rural area near your home you are stopped by a county police officer who informs you that you have been clocked (with radar) at 38 miles per hour in a 25-mph zone. You believe this information to be accurate. After the policeman leaves, you inspect your citation and find that the details on the summons regarding weather, visibility, time, and location of violation are highly inaccurate. The citation informs you that you may either \textbf{pay a \$20 fine} by mail without appearing in court or you must appear in municipal court within the next two weeks to \textbf{contest the charge}.

        \dotfill\\
        
        What should you do, \textbf{Pay speeding fine} or \textbf{Contest charge}?
    \end{boxedminipage}
    \begin{boxedminipage}[c][18em][t]{\columnwidth}
        \textbf{Story 4. Space R\&D program}\\
        
        \small It is proposed in Congress that the space program be revived and that large sums be allocated for the manned and unmanned exploration of the moon and planets nearest Earth. Supporters of the proposal argue that it will provide jobs. spur technology, and promote national pride and unity. Opponents argue that a space program will either necessitate higher taxes, or else dram money from important domestic priorities. Furthermore, they deny that it will accomplish the desirable effects claimed by the program’s supporters. Both sides, of course, refute each other’s claims and ultimately a public referendum is held.

        \dotfill\\
        
        What should you do, \textbf{Vote for cutback} or \textbf{Vote against cutback}?
    \end{boxedminipage}
    \caption{Four stories and queries used in \citet{ross1977false}}
    \label{fig:stories}
\end{figure*}

\section{Environment for Experiment}
\label{app:env}
Here, we briefly illustrate the environment used for our experiment.

\subsection{Study 1}
All the experiments were done in the following environment. For the hardware system, we used a Macbook Pro with an Apple M3 Pro chip. For the software system, the system has MacOS Sonoma 14.1 with Python 3.10.13. We also used Python libraries including \texttt{openai} 0.28.0, \texttt{groq} 0.4.2, \texttt{anthropic} 0.21.1, \texttt{pandas} 2.1.4, \texttt{statsmodels} 0.14.0, and \texttt{scipy} 1.11.4 for the experiment.

\subsection{Study 2}
The experimental setup is the same as that of Study 1, except for the prompting styles. We tested the same LLMs, GPT-4, Claude 3, LLaMA 2, and Mixtral, as in Study 1. All the experiments were conducted from February 24, 2024, to June 10, 2024, by calling 5120 API calls\footnote{40 characters, 4 stories, 2 options, 16 condition pairs.} for each LLM\footnote{Code: [anonymized for the review]}.

\section{Detailed result for Study 1}
\label{app:study1-result}

\subsection{Shapiro-Wilk normality test}
\label{app:study1-normality}
Before conducting a statistical test, we checked whether our experimental result followed a normal distribution using the Shapiro-Wilk test. The null hypothesis of this test is that 'the data follows a normal distribution.' Thus, a p-value under 0.05 indicates that the provided data is not normal. Table \ref{tab:normality} (page \pageref{tab:normality}) shows the result. As shown, more than half of our experimental results are not normal. Therefore, using a t-test or an ANOVA is not suitable because they assume normality. Thus, we chose to use non-parametric analyses in further statistical analyses.

\input{table/normality}

\subsection{Comparing LLMs with human (H1-1)}
\label{app:study1-h11}

Though we aim to adopt human experiments to understand LLMs' behavior, one can ask whether the tendency is similar to humans. As we borrowed the experimental design from \citet{ross1977false} and \citet{cross-culture}, we can compare our results on LLMs with previous reports on human FCE. Table \ref{tab:human} shows the result of human FCE with our experimental result. Note that the last four rows are the same as Table \ref{tab:h1-fce}, as we copied the data from that Table. Note that a human study used a parametric test (t-test) to identify FCE.

\input{table/human-compare}

\subsection{Result of free-form answers}
\label{app:free-form}

We also report the results of the H1-1 test on free-form answers. In the main manuscript, we reported the result when we directly fed each option as if LLMs had chosen that option. However, this may not reflect the actual tendency of LLM's generation procedure, because we prevent LLMs from generating their preferred options. Therefore, to fill the gap between real-world scenarios and our experimental setting, we decided to provide the result of such free-form answers. As the statistical test requires perceived agreements of both options, we used combinations where LLMs chose both answers. As shown in Table \ref{tab:own-choice}, there were four combinations: Mixtral on Story 1, Claude 3 on Story 3, and GPT-4/ LLaMA 2 on Story 4.

The result of the free-form setting is shown in Table \ref{tab:free-form}. As shown, we observed FCE even in the free-form setting. Thus, we concluded that the result is not much different from the result we reported in Table \ref{tab:h1-fce}.

\input{table/freeform}

\subsection{Detailed result for H1-2 (Culture)}
\label{app:study1-h12}

Tables \ref{tab:h1-american} and \ref{tab:h1-korean} (page \pageref{tab:h1-korean}) show the detailed results for each group, European American and Korean, respectively. Each table shows the Mann-Whitney U test result to provide insight into how FCE occurred in each persona group.

\input{table/h1-america}
\input{table/h1-korea}

\subsection{Detailed result for H1-3 (Gender)}
\label{app:study1-h13}

Tables \ref{tab:h1-male} and \ref{tab:h1-female} (page \pageref{tab:h1-female}) show the detailed results for each group, Male and Female. Each table shows the Mann-Whitney U test result to provide insight into how FCE occurred in each persona group.

\input{table/h1-man}
\input{table/h1-woman}

\section{Prompt design of Study 2}
\label{app:study2-design}

To experiment with Study 2, we (1) added reasoning information to the user prompts and (2) adopted deep reasoning methods. Basically, the structure of the input prompt is the same as Study 1 until we ask LLMs to estimate the perceived agreement of other peers. First, for additional reasoning information, we used the following prompts to deliver reasoning of LLM's choice before asking about perceived agreement. See Section \ref{app:info} for the detailed reasoning input. Second, for deep reasoning methods, we modified the last question to ask direct reasoning for an LLM's estimation. See Section \ref{app:chain} for the details.

\begin{boxedminipage}{.9\columnwidth}\small
\textbf{Assistant's answer}: (Same as Study 1)\\
(Instead of generation, we feed the following)
\begin{verbatim}
[Selected option]
\end{verbatim} 

\textbf{Assistant's answer}: (Feed reasoning)\\
\begin{verbatim}
[Reasoning for the choice]
\end{verbatim}

And then, experimenter will ask perceived agreement, using methods in reasoning chain. By default, we use condition R1, as follows:
\textbf{User prompt}:
\begin{verbatim}
What % of your peers do you estimate would 
 each option?(Total % should be 100%).
\end{verbatim}
\end{boxedminipage}

\subsection{Provided information}
\label{app:info}

For simplicity, we focus on the effect of reasoning text generated when LLM makes its own choice. So, we simulate LLM's own reasoning in P2-P4. The following subsections describe the actual reasoning generated by each LLM and used in this study. We used them as supportive or opposite information in testing H2-1 of Study 2. For example, if we are simulating the assistant's answer as option 1, we use the reasoning of option 1 as supportive and option 2 as the opposite. For the irrelevant reasoning, we used the following paragraph.

\begin{itemize}\itemsep0pt\small
    \item \textbf{Irrelevant information (68 tokens)} : \texttt{Under the twinkling starlight, colorful unicorns dance vigorously against the mysterious night sky, riding a magical rainbow across the Milky Way to the best ice cream party in the universe. At the party, unicorns create and share their own unique flavors and colors of ice cream, spending a laughable time with friends from all over the galaxy.}
\end{itemize}

\subsubsection{GPT-4}
\paragraph{Reasoning used for Story 1}
\begin{itemize}\itemsep0pt\small
    \item \textbf{Option 1 (68 tokens)}: \texttt{I chose 'individual paper' because it allows you to have full control over the quality and outcome of your work. In a group setting, there's a risk of unequal effort distribution, potentially impacting your grade negatively. Individual work ensures that your grade reflects your own effort and understanding, providing a clear measure of your mastery of the course material.}
    \item \textbf{Option 2 (67 tokens)}: \texttt{I chose the group paper option because it fosters collaboration, allowing team members to pool their knowledge and skills, leading to a more comprehensive and diverse paper. It also offers an opportunity to learn from peers, develop teamwork skills, and share the workload, potentially reducing the individual stress and time commitment that comes with completing a large project alone.}
\end{itemize}

\paragraph{Reasoning used for Story 2}
\begin{itemize}\itemsep0pt\small
\item \textbf{Option 1 (68 tokens)} :  \texttt{I chose 'Sign release' because if you genuinely like the store and your comments were sincere, allowing them to use your feedback can help the store you appreciate and possibly contribute positively to your community. Plus, it can be a unique experience to see yourself contributing to a local business's marketing efforts, fostering a sense of community and personal connection.} \\
\item \textbf{Option 2 (66 tokens)} : \texttt{I chose 'Not sign release' because your consent was not obtained before recording, raising ethical concerns. Signing the release after being unknowingly filmed can set a precedent for privacy infringement. Moreover, without control over how your comments are edited or used, you risk misrepresentation or unintended endorsement, which could impact your personal image or privacy.}
\end{itemize}

\paragraph{Reasoning used for Story 3}
\begin{itemize}\itemsep0pt\small
\item \textbf{Option 1 (70 tokens)}: \texttt{I chose 'Pay speeding fine' because contesting the charge would likely require time, effort, and potentially additional costs that might outweigh the \$20 fine. Given that the primary violation (speeding) is admitted to be accurate, contesting the inaccuracies in details may not significantly impact the outcome but could lead to unnecessary legal complexities and personal inconvenience.} \\
\item \textbf{Option 2 (68 tokens)}: \texttt{I chose 'Contest charge' because the inaccuracies on the citation regarding weather, visibility, time, and location suggest a lack of attention to detail, which could undermine the credibility of the speeding charge. Contesting provides an opportunity to address these discrepancies, possibly leading to a dismissal or reduction of the charge, despite acknowledging the speeding violation.}
\end{itemize}

\paragraph{Reasoning used for Story 4}
\begin{itemize}\itemsep0pt\small
\item \textbf{Option 1 (65 tokens)}: \texttt{I chose 'Vote for cutback' because, despite the potential benefits of a space program like job creation, technological advancement, and national pride, the significant financial costs and potential diversion of funds from essential domestic priorities cannot be overlooked. In a scenario where resources are finite, prioritizing immediate societal needs and fiscal responsibility is crucial.} \\
\item \textbf{Option 2 (68 tokens)}: \texttt{I chose to vote against the cutback because investing in space exploration can drive technological advancements, create employment opportunities, enhance national pride, and potentially yield long-term benefits that outweigh the initial costs. The potential for significant scientific discoveries and the broader economic and educational impacts justify prioritizing space exploration over immediate concerns, despite the valid points raised by opponents.}
\end{itemize}

\subsubsection{Cluade3}
\paragraph{Reasoning used for Story 1}
\begin{itemize}\itemsep0pt\small
\item \textbf{Option 1 (67 tokens)}: \texttt{I would choose an individual paper to have full control over my work and grade. Relying on others for a major paper is risky, as their performance directly impacts my grade. Working individually allows me to manage my time, ensure the quality of my work, and be graded based on my own effort and understanding of the course material.} \\
\item \textbf{Option 2 (68 tokens)}: \texttt{Choosing a group paper allows for collaboration, sharing of ideas, and division of work, potentially leading to a higher quality paper. It also provides a safety net, as the risk of a low grade is spread among the team. However, individual work allows for full control and recognition of one's efforts, which may be preferable for some students.}
\end{itemize}

\paragraph{Reasoning used for Story 2}
\begin{itemize}\itemsep0pt\small
\item \textbf{Option 1 (66 tokens)}: \texttt{I would sign the release because I genuinely enjoy shopping at the supermarket and have no issue with sharing my honest opinion. The commercial could help promote a local business that I support, and my comments were truthful and positive. As long as the footage is unedited and used appropriately, I see no harm in allowing them to use it.} \\
\item \textbf{Option 2 (67 tokens)}: \texttt{I would not sign the release because I value my privacy and do not want my unplanned, candid comments used in a commercial without my informed consent. The supermarket should have been transparent about their intentions from the start. I would politely decline and suggest that they find willing participants who can provide a more thoughtful, prepared endorsement of their store.}
\end{itemize}

\paragraph{Reasoning used for Story 3}
\begin{itemize}\itemsep0pt\small
\item \textbf{Option 1 (66 tokens)}: \texttt{Paying the fine is the most practical choice. The inaccuracies in the citation details are minor and unlikely to affect the outcome of the case. Contesting the charge would require taking time off work to appear in court, with no guarantee of success. Paying the \$20 fine by mail is the simplest and most efficient resolution.} \\
\item \textbf{Option 2 (67 tokens)}: \texttt{Contest the charge because the citation contains significant inaccuracies about the details of the incident. While you acknowledge exceeding the speed limit, the officer's report is flawed. Appearing in court allows you to present your case, highlight the discrepancies, and potentially have the charge reduced or dismissed due to the officer's errors in documenting the violation.}
\end{itemize}

\paragraph{Reasoning used for Story 4}
\begin{itemize}\itemsep0pt\small
\item \textbf{Option 1 (66 tokens)}: \texttt{I would vote for a cutback in the space program. While space exploration has merits, allocating large sums to it would strain the budget and divert funds from pressing domestic issues. The claimed benefits of jobs, technology, and unity are uncertain. Given current priorities, it's more prudent to focus resources on immediate needs here on Earth.} \\
\item \textbf{Option 2 (68 tokens)}: \texttt{I would vote against the cutback because investing in space exploration can drive technological innovation, create jobs, and inspire future generations. While there are important domestic priorities, a well-funded space program can provide long-term benefits to society. It's crucial to balance spending, but cutting back on space exploration may hinder scientific progress and limit potential economic opportunities.}
\end{itemize}

\subsubsection{LLaMA2}
\paragraph{Reasoning used for Story 1}
\begin{itemize}\itemsep0pt\small
\item \textbf{Option 1 (66 tokens)}: \texttt{I chose individual paper because I prefer to have full control over my work. This ensures that time available brings a desirability that adds flexibility and superior comprehension with work deliverables, including case study scenarios and essays written collaboratively but individually accounted for a higher academic assessment. This, in turn makes proper citations advised for personal success.} \\
\item \textbf{Option 2 (66 tokens)}: \texttt{I chose group paper because it allows me to collaborate with others, share ideas and learn from one another. It also distributes the workload and provides an opportunity to develop teamwork skills, which are essential in today's workplace. Additionally, I believe that the diverse perspectives and skills of three people can create a more comprehensive and well-rounded paper.}
\end{itemize}

\paragraph{Reasoning used for Story 2}
\begin{itemize}\itemsep0pt\small
\item \textbf{Option 1 (68 tokens)}: \texttt{I chose to sign the release because the man in the business suit revealed that a videotape crew filmed my comments, and it's an opportunity for me to be featured in a TV commercial for the supermarket chain, which could be a fun and unique experience. Additionally, it would be a way to support the supermarket that I frequent and appreciate.} \\
\item \textbf{Option 2 (66 tokens)}: \texttt{I chose not to sign the release because I don't want my image and comments to be used for commercial purposes without my full understanding of how they will be used and potentially taken out of context. Additionally, I haven't been provided with enough information about the context in which the footage will be used and how it will be edited..}
\end{itemize}

\paragraph{Reasoning used for Story 3}
\begin{itemize}\itemsep0pt\small
\item \textbf{Option 1 (68 tokens)}: \texttt{I chose to pay the speeding fine because I believe it's the most convenient and efficient option. It saves time and effort compared to contesting the charge in court, which may require multiple appearances and potentially lengthy proceedings. Additionally, paying the fine allows me to resolve the issue quickly and move on without the stress and uncertainty of a court case.} \\
\item \textbf{Option 2 (69 tokens)}: \texttt{I chose to contest the charge because the details on the summons regarding weather, visibility, time, and location of violation are highly inaccurate. I believe that these inaccuracies may indicate a faulty speed-measuring device or other error on the part of the policeman, and I would like to bring this to the attention of the court and clear my name.}
\end{itemize}

\paragraph{Reasoning used for Story 4}
\begin{itemize}\itemsep0pt\small
\item \textbf{Option 1 (67 tokens)}: \texttt{I chose to vote for cutback because I believe that the funds allocated for the space program could be better used to address more pressing domestic issues, such as poverty, healthcare, and education. Additionally, I think that the benefits of space exploration are not sufficient to justify the high costs, especially when there are more urgent needs on Earth.} \\
\item \textbf{Option 2 (67 tokens)}: \texttt{I chose to vote against cutback because I believe that investing in the space program can have long-term benefits for society, such as advancing technology and promoting national pride and unity. Additionally, it could lead to new discoveries and opportunities for humanity. While it may require significant funding upfront, the potential benefits outweigh the costs in the long run.}
\end{itemize}

\subsubsection{Mixtral}
\paragraph{Reasoning used for Story 1}
\begin{itemize}\itemsep0pt\small
\item \textbf{Option 1 (68 tokens)}: \texttt{I chose individual papers because it provides a better opportunity to demonstrate my own knowledge and abilities, and I believe it will result in a more accurate and fair assessment of my performance. Additionally, working individually allows for greater flexibility in choosing a topic and schedule, as well as the development of important skills such as self-motivation and time management.} \\
\item \textbf{Option 2 (68 tokens)}: \texttt{I chose group paper because it allows for collaboration and diverse perspectives, which can lead to a higher quality paper. Additionally, group work can help build teamwork and problem-solving skills, which are valuable in the major area of study. The risk of receiving a lower grade is shared among the group, and the potential for learning and growth is increased.}
\end{itemize}

\paragraph{Reasoning used for Story 2}
\begin{itemize}\itemsep0pt\small
\item \textbf{Option 1 (69 tokens)}: \texttt{I chose to sign the release because the supermarket in question genuinely offers high-quality meat and produce at reasonable prices, and I stand by my positive review of the store. By signing the release, I allow the supermarket to use my comments in their TV commercial, potentially encouraging others in my community to shop there as well, and supporting a local business.} \\
\item \textbf{Option 2 (68 tokens)}: \texttt{I chose not to sign the release because the situation was unexpected and unsolicited. The filming was not previously disclosed, and the request for a release was made only after the footage had already been taken. It's important to maintain control over one's image and personal information, and signing a release without proper consideration could potentially lead to unintended consequences.}
\end{itemize}

\paragraph{Reasoning used for Story 3}
\begin{itemize}\itemsep0pt\small
\item \textbf{Option 1 (68 tokens)}: \texttt{I chose to pay the speeding fine because the radar evidence for the speed violation is likely accurate, while the other details on the citation are inaccurate but not material to the charge. Contesting the charge may result in additional fines, time, and effort, and the outcome is uncertain. Paying the fine is a quicker and more straightforward resolution.} \\
\item \textbf{Option 2 (69 tokens)}: \texttt{I recommended contesting the charge because the citation contains inaccuracies regarding weather, visibility, time, and location of the violation. These inaccuracies could potentially weaken the case against the driver, and increase the chances of having the charge reduced or dismissed. It's important to ensure that all information on a citation is accurate before deciding to pay a fine.}
\end{itemize}

\paragraph{Reasoning used for Story 4}
\begin{itemize}\itemsep0pt\small
\item \textbf{Option 1 (66 tokens)}: \texttt{I chose to vote for cutbacks in the space program due to concerns about the potential impact on taxes and the opportunity cost of funding space exploration at the expense of other domestic priorities. I believe that the benefits of a space program may not outweigh the costs, particularly if it results in higher taxes or reduced funding for important domestic programs.} \\
\item \textbf{Option 2 (67 tokens)}: \texttt{I chose to vote against cutbacks in space exploration because it has the potential to significantly benefit society through job creation, technology advancement, and national pride. Additionally, it can inspire future generations to pursue careers in STEM fields, and the discoveries made through space exploration could have far-reaching implications for our understanding of the universe and our place in it.}
\end{itemize}

\subsection{Reasoning chain}
\label{app:chain}

\subsubsection{R1: Direct condition}
When asking about the perceived agreement, we used the following prompt:

\begin{boxedminipage}{.9\columnwidth}\small
\textbf{User prompt}:
\begin{verbatim}
What % of your peers do you estimate would 
 each option?(Total % should be 100%).
Just tell your answer without any reasoning
\end{verbatim}
\end{boxedminipage}

\subsubsection{R2: Simple condition}
When asking about the perceived agreement, we used the following prompt:

\begin{boxedminipage}{.9\columnwidth}\small
\textbf{User prompt}:
\begin{verbatim}
What % of your peers do you estimate would 
 each option?(Total % should be 100%).
Tell your answer and provide your reasoning.
\end{verbatim}
\end{boxedminipage}

\subsubsection{R3: CoT condition}
When asking about perceived agreement, we let the model generate reasoning first and then generate the answer using the following prompts.

\begin{boxedminipage}{.9\columnwidth}\small
\textbf{User prompt}:
\begin{verbatim}
What % of your peers do you estimate would 
each option? (Total % should be 100%). 
Just tell your answer without any reasoning.
\end{verbatim}
\textbf{Assistant's answer} (Generated):
\begin{verbatim}
    [Answer]
\end{verbatim}
\textbf{User prompt}: after generating
\begin{verbatim}
Let's think step by step. 
What % of your peers do you estimate would
 each option?(Total % should be 100%).
\end{verbatim}

\textbf{Assistant's answer} (Generated):
\begin{verbatim}
[Reasoning]
\end{verbatim}

\end{boxedminipage}

\subsubsection{R4: Reflection condition}
When asking for perceived agreement, we let the model generate reasoning first, rethink their reasoning, and generate the answer using the following prompts.

\begin{boxedminipage}{.9\columnwidth}\small
\textbf{User prompt}:
\begin{verbatim}
Let's think step by step. 
What % of your peers do you estimate would 
 each option?(Total % should be 100%).
\end{verbatim}

\textbf{Assistant's answer} (Generated):
\begin{verbatim}
[Reasoning]
\end{verbatim}

\textbf{User prompt}, after rethinking:
\begin{verbatim}
Consider the given situation again and read
 your reasoning according to the given 
 situation. If required, rewrite your 
 reasoning by applying necessary changes to
 improve your prediction.
What % of your peers do you estimate would 
 each option?(Total % should be 100%). 
Tell your answer and provide your reasoning.
\end{verbatim}
\end{boxedminipage}

\section{Detailed result for Study 2}
\label{app:study2-result}

\subsection{Detailed result for H2-1 (Info)}
\label{app:study2-h21}
In this section, we present the results for H2-1 (Provided information). The Kruskal-Wallis Test results for H2-1 are at the top of Table \ref{tab:h2-summary} and Table \ref{tab:h2-summary-app}. Additionally, detailed experimental results for H2-1 can be found in Table \ref{tab:h2-1-app-story1}, Table \ref{tab:h2-1-app-story2}, Table \ref{tab:h2-1-app-story3}, and Table \ref{tab:h2-1-app-story4}.

\paragraph{Story 1} Significant differences are noted for GPT-4 between P1 - P4, P2 - P4, and P3 - P4. For Claude3, significant differences are present between P1 - P3, P1 - P4, P2 - P3, and P2 - P4. LLaMA2 shows significant differences between P1 - P2, P1 - P3, and P1 - P4. Mixtral exhibits significant differences between P1 - P3, P2 - P3, P2 - P4, and P3 - P4.

\paragraph{Story 2} For GPT-4, significant differences are noted in all scenarios except P2 - P4. For Claude3, significant differences are present between P1 - P3, P2 - P3, and P3 - P4. For LLaMA2, significant differences are noted in all scenarios except P1 - P4. For Mixtral, significant differences exist between P1 - P3, P2 - P3, and P3 - P4.

\paragraph{Story 3} For GPT-4, significant differences are noted in all scenarios except P2 - P4. For Claude3, significant differences were not observed in any of the stories. For LLaMA2, significant differences are present between P1 - P2, P2 - P3, and P2 - P4. For Mixtral, significant differences exist between P1 - P3, P2 - P3, and P3 - P4.

\paragraph{Story 4} For GPT-4, significant differences exist between P1 - P3, P2 - P3, and P3 - P4. For Claude3, significant differences are present between P1 - P3, P1 - P4, P2 - P4, and P3 - P4. For LLaMA2, significant differences are present between P1 - P3, P2 - P3, P2 - P4, and P3 - P4. For Mixtral, significant differences exist between P1 - P2, P1 - P3, P1 - P4, and P3 - P4.

\subsection{Range of LLMs estimation}
\label{app:study2-range}
In this section, we provide a detailed explanation of the range of LLM's estimation. Table \ref{tab:h2-range} shows the percentage of answers in a particular range. As shown in the Table, most LLMs' answers were between 20\% and 80\%.

\subsection{Detailed result for H2-2 (Chain)}
\label{app:study2-h22}
In this section, we provide a detailed explanation of the results for H2-2 (Chain). The results of the Kruskal-Wallis Test for H2-2 are observed at the bottom of Tables \ref{tab:h2-summary} and \ref{tab:h2-summary-app}. Additionally, detailed experimental results for H2-2 can be found in Tables \ref{tab:h2-2-app-story1}, \ref{tab:h2-2-app-story2}, \ref{tab:h2-2-app-story3}, and \ref{tab:h2-2-app-story4}.

\paragraph{Story 1}
For GPT-4, significant differences are noted between R1 - R3, R2 - R3, and P3 - P4. For Claude3, significant differences are pointed out in all stories. LLaMA2 shows significant differences between R1 - R4. Mixtral exhibits significant differences between R1 - R2 and R1 - P3.

\paragraph{Story 2}
For GPT-4, significant differences exist between R1 - R4, R2 - R4, and R3 - R4. For Claude3, significant differences are present between R1 - R3, R1 - R4, R2 - R3, and R2 - R4. For LLaMA2, significant differences were not observed in any of the stories. For Mixtral, significant differences are noted in all scenarios except R1 - R4.

\paragraph{Story 3}
For GPT-4, significant differences are present between R1 - R2, R1 - R4, R2 - R3 and R2 - R4. For Claude3, significant differences are noted in all scenarios except R1 - R2. For LLaMA2, significant differences are present between R1 - R4, R2 - R4, and R3 - R4. For Mixtral, significant differences exist between R1 - R4, R2 - R4, and R3 - R4.

\paragraph{Story 4}
For GPT-4, significant differences exist between R1 - R4, R2 - R4, and R3 - R4. For Claude3, significant differences are noted in all stories.
For LLaMA2, significant differences are present between R1 - R2, R1 - R3, R1 - R4 and R2 - R4. For Mixtral, significant differences were not observed in any of the stories.

\subsection{Exploration of the interaction effect}
\label{app:study2-int}

\begin{figure*}[p]
    \centering
    \includegraphics[height=.1833\textheight]{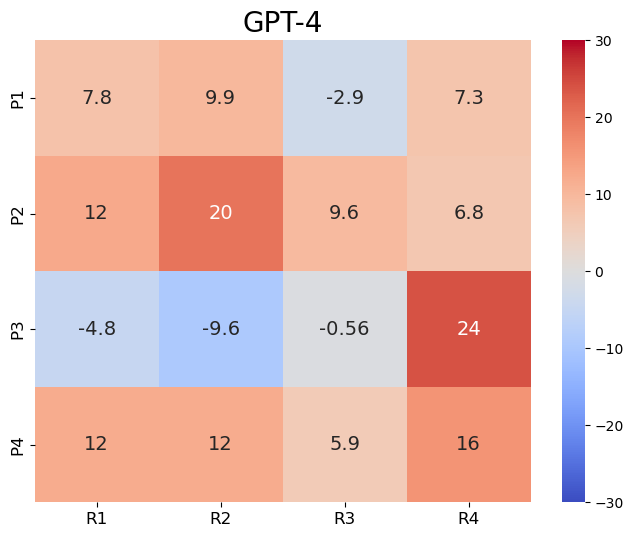} \includegraphics[height=.1833\textheight]{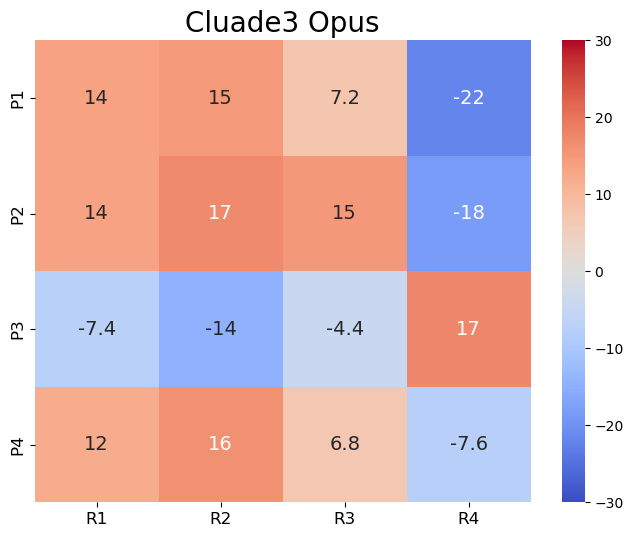}

    \includegraphics[height=.1833\textheight]{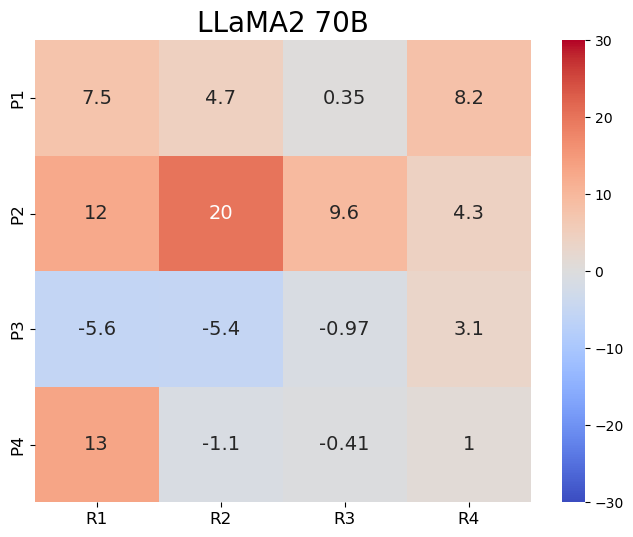}    \includegraphics[height=.1833\textheight]{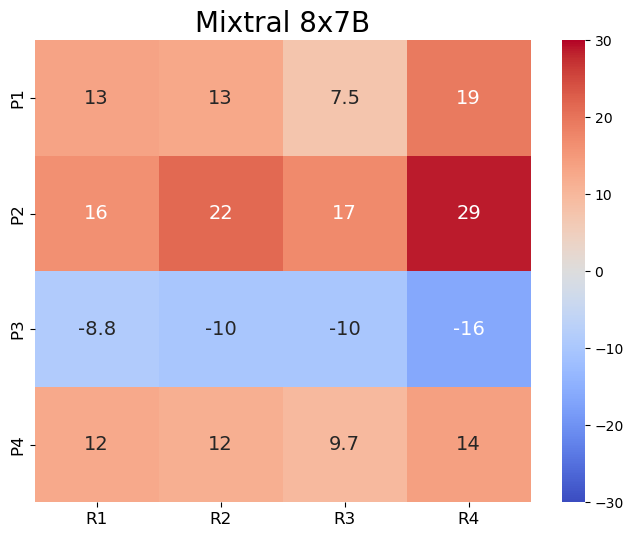}
    \caption{Heatmap of FCE strength, representing interaction effect between two prompting styles, for each model.}
    \label{fig:heatmap}
\end{figure*}

In this section, We draw heatmap images to examine the tendency of the interaction effect. Figure \ref{fig:heatmap} shows the heatmap for each model.

\paragraph{GPT-4} FCE was strongest with (P3, R4), while (P3, R3) yielded approximately neutral responses. Meanwhile, with (P3, R2), GPT-4 provided answers that were most strongly opposite to its own reasoning, making the FCE weakest.

\paragraph{Claude 3} FCE was strongest with (P3, R4) and (P2, R2), while (P3, R3) yielded approximately neutral responses. This result is similar to GPT-4. Meanwhile, Claude 3 provided answers most strongly opposite to its own reasoning with (P1, R4), making the FCE weakest. Still, (P3, R2) showed negative strength in FCE, as in GPT-4.

\paragraph{LLaMA 2} FCE was strongest with (P2, R2), while (P1, R3) or (P3, R3) yielded approximately neutral responses. This result is somewhat similar to Claude 3. Meanwhile, LLaMA 2 provided answers most strongly opposite to its own reasoning with (P3, R1), making the FCE weakest. Similar to Claude 3, (P3, R2) still showed negative strength in FCE.

\paragraph{Mixtral} FCE was strongest with (P2, R4), followed by (P2, R2). This is similar to Claude 3 and LLaMA 2. For the neutral responses, there is no condition whose strength is near zero. However, similar to other models, (P1, R3) showed the lowest absolute value in FCE strength. Meanwhile, Mixtral provided answers most strongly opposite to its own reasoning with (P3, R4), making the FCE weakest. Similar to other models, (P3, R2) still showed negative strength in FCE.

\input{table/range}

\input{table/h2_appendix}

\input{table/story1}

\input{table/story2}

\input{table/story3}

\input{table/story4}

\end{document}

%% file: table/skew.tex
\begin{table}
    \centering
    \small
    \begin{tabular}{r|rr|rr}
    \toprule
            & \multicolumn{2}{c}{Option in \textbf{Story 1}} & \multicolumn{2}{c}{Options in \textbf{Story 2}} \\
        \cmidrule(lr){2-3}\cmidrule(lr){4-5}
        & Individual & Group & Sign & Not sign \\
    \midrule
    GPT-4    & 40 & 0 & 40 &  0\\
    Claude 3 & 40 & 0 &  0 & 40\\
    LLaMA 2  & 40 & 0 & 40 &  0\\
    Mixtral  & 37 & 3 &  0 & 40\\
    \toprule
            & \multicolumn{2}{c}{Option in \textbf{Story 3}} & \multicolumn{2}{c}{Options in \textbf{Story 4}} \\
        \cmidrule(lr){2-3}\cmidrule(lr){4-5}
        & Pay fine & Contest & Vote for & against \\
    \midrule
    GPT-4    & 0 & 40 & 9 & 31\\
    Claude 3 & 20 & 20 & 40 & 0\\
    LLaMA 2  & 0 & 40 & 31 & 9\\
    Mixtral  & 0 & 40 & 0 & 40\\
    \bottomrule 
    \end{tabular}
    \caption{Skewness of LLMs answer. A cell shows the number of choosing options 1 and 2 in each story.}
    \label{tab:own-choice}
\end{table}


%% file: table/h1-1.tex
\begin{table}[!t]
    \centering
    \small
    \begin{tabular}{@{}lcc|cc@{}l@{}}
        \toprule
         \multicolumn{6}{l}{\textbf{Story 1. Term paper:}  Individual \textit{vs.} Group} \\
         \midrule
         & \multicolumn{2}{c|}{Perceived Agreement on} & \multicolumn{3}{c}{Mann-Whitney}\\
         \cmidrule{2-3}\cmidrule{4-6}
         & Individual & Group & Diff. & Stat. &\\
        \midrule
        GPT-4
        & 60.0 & 59.7 &  0.3 & 820 &  \\
        
        Claude 3
        & 60.0 & 40.0 & 20.0 & 1600 & \starxxx \\

        LLaMA 2
        & 60.3 & 49.0 & 11.3 & 1249 & \starxxx \\
        
        Mixtral
        & 60.9 & 60.0 &  0.9 & 901 & \\

        \midrule

        Human 
        & 67.4 & 45.9 & 21.5 & & \starxxx\\
        
                \toprule
         \multicolumn{6}{l}{\textbf{Story 2. Supermarket:}  Sign \textit{vs.} Not sign} \\
         \midrule
         & \multicolumn{2}{c|}{Perceived Agreement on} & \multicolumn{3}{c}{Mann-Whitney}\\
         \cmidrule{2-3}\cmidrule{4-6}
         & Sign & Not sign & Diff. & Stat. &\\
        \midrule
        GPT-4
        & 60.3 & 52.0 &  8.3 & 1332 & \starxxx \\
        
        Claude 3
        & 61.0 & 35.5 & 25.5 & 1600 & \starxxx \\

        LLaMA 2
        & 70.0 & 69.0 & 1.0 & 880 & \starx \\
        
        Mixtral
        & 76.3 & 35.8 &  40.5 & 1577 & \starxxx\\

        \midrule

        Human & 
        75.6 & 57.3 & 18.3 & & \starxxx\\
        
        \toprule
         \multicolumn{6}{l}{\textbf{Story 3. Traffic Ticket:}  Pay fine \textit{vs.} Contest} \\
         \midrule
         & \multicolumn{2}{c|}{Perceived Agreement on} & \multicolumn{3}{c}{Mann-Whitney}\\
         \cmidrule{2-3}\cmidrule{4-6}
         & Pay fine & Contest & Diff. & Stat. &\\
        \midrule
        GPT-4
        & 62.5 & 60.0 &  2.5 & 1000 & \starxxx \\
        
        Claude 3
        & 70.0 & 70.0 & 0.0 & 800 &  \\

        LLaMA 2
        & 70.0 & 70.0 & 0.0 & 880 &  \\
        
        Mixtral
        & 65.8 & 56.5 & 9.3 & 1233 & \starxxx \\

        \midrule

        Human 
        & 71.8 & 51.7 & 20.1 & & \starxxx \\

        \toprule
         \multicolumn{6}{l}{\textbf{Story 4. Space R\&D:}  Vote for \textit{vs.} Vote against} \\
         \midrule
         & \multicolumn{2}{c|}{Perceived Agreement on} & \multicolumn{3}{c}{Mann-Whitney}\\
         \cmidrule{2-3}\cmidrule{4-6}
         & Vote for & Vote against & Diff. & Stat. &\\
        \midrule
        GPT-4
        & 60.0 & 40.0 & 20.0 & 1600 & \starxxx \\
        
        Claude 3
        & 60.0 & 50.9 & 9.1 & 1160 & \starxxx \\

        LLaMA 2
        & 61.3 & 43.5 & 17.8 & 1477 & \starxxx \\
        
        Mixtral
        & 42.9 & 39.6 & 3.3 & 1095 & \starxxx \\

        \midrule

        Human 
        & 65.7 & 48.5 & 17.2 & & \starx \\

        \toprule
        \multicolumn{6}{r}{{\small \starx $p < 0.05$, \starxx $p < 0.01$, \starxxx $p < 0.001$}}
    \end{tabular}
    \caption{Mann-Whitney U Test for H1-1. \textit{Diff.} and \textit{Stat.} columns indicates the agreement difference between two options and $U$-statistics from the test, respectively. Human results are copied from \citep{ross1977false}.}
    \label{tab:h1-fce}
\end{table}

%% file: table/h1-2.tex
\begin{table}[!t]
    \centering
    \small
    \begin{tabular}{lcc|cc@{}l}
        \toprule
         \multicolumn{6}{l}{\textbf{Story 1. Term paper:}  Individual \textit{vs.} Group} \\
         \midrule
         & \multicolumn{2}{c|}{FCE per group} & \multicolumn{3}{c}{Kruskal-Wallis}\\
         \cmidrule{2-3}\cmidrule{4-6}
         & Korean & American & Diff. & Stat. &\\
        \midrule
        GPT-4
        & 0.0 & 0.5 & -0.5 & 1.0 &  \\
        
        Claude 3
        & 20.0 & 20.0 & 0.0 & / &  \\

        LLaMA 2
        & 17.5 & 5.0 & 12.5 & 14.4 & \starxxx \\
        
        Mixtral
        & 1.0 & 0.8 &  0.2 & 0.0 & \\

                \toprule
         \multicolumn{6}{l}{\textbf{Story 2. Supermarket paper:}  Sign \textit{vs.} No sign} \\
         \midrule
         & \multicolumn{2}{c|}{per group} & \multicolumn{3}{c}{Kruskal-Wallis}\\
         \cmidrule{2-3}\cmidrule{4-6}
         & Korean & American & Diff. & Stat. &\\
        \midrule
        GPT-4
        & 15.5 & 1.0 & 14.5 & 21.9 & \starxxx \\
        
        Claude 3
        & 31.0 & 20.0 & 11.1 & 33.4 & \starxxx \\

        LLaMA 2
        & 0.0 & 2.0 & -2.0 & 4.3 & \starx \\
        
        Mixtral
        & 41.0 & 40.0 & 1.0 & 0.2 &  \\

        \toprule
         \multicolumn{6}{l}{\textbf{Story 3. Traffic Ticket:}  Pay fine \textit{vs.} Contest} \\
         \midrule
         & \multicolumn{2}{c|}{FCE per group} & \multicolumn{3}{c}{Kruskal-Wallis}\\
         \cmidrule{2-3}\cmidrule{4-6}
         & Korean & American & Diff. & Stat. &\\
        \midrule
        GPT-4
        & 2.0 & 3.0 & -1.0 & 0.5 & \\
        
        Claude 3
        & 0.0 & 0.0 & 0.0 & / &  \\

        LLaMA 2
        & 0.0 & 0.0 & 0.0 & / &  \\
        
        Mixtral
        & 8.5 & 10.0 & -1.5 & 0.3 &  \\

        \toprule
         \multicolumn{6}{l}{\textbf{Story 4. Space R\&D:}  Vote for \textit{vs.} Vote against} \\
         \midrule
         & \multicolumn{2}{c|}{FCE per group} & \multicolumn{3}{c}{Kruskal-Wallis}\\
         \cmidrule{2-3}\cmidrule{4-6}
         & Korean & American & Diff. & Stat. &\\
        \midrule
        GPT-4
        & 20.0 & 20.0 & 0.0 & / &  \\
        
        Claude 3
        & 1.0 & 17.3 & -16.3 & 24.9 & \starxxx \\

        LLaMA 2
        & 16.5 & 19.0 & -2.5 & 0.5 &  \\
        
        Mixtral
        & 2.0 & 4.5 & 2.5 & 2.8 &  \\

        \toprule
        \multicolumn{6}{r}{{\small \starx $p < 0.05$, \starxx $p < 0.01$, \starxxx $p < 0.001$}}
    \end{tabular}
    \caption{Kruskal-Wallis test for H1-2 (Culture). \textit{Diff.} and \textit{Stat.} columns indicates agreement difference between two options and $H$-statistics from the test, respectively.}
    \label{tab:h1-2-summary}
\end{table}

\begin{table}[!t]
    \centering
    \small
    \begin{tabular}{lcc|cc@{}l}
        \toprule
         \multicolumn{6}{l}{\textbf{Story 1. Term paper:}  Individual \textit{vs.} Group} \\
         \midrule
         & \multicolumn{2}{c|}{FCE per group} & \multicolumn{3}{c}{Kruskal-Wallis}\\
         \cmidrule{2-3}\cmidrule{4-6}
         & Male & Female & Diff. & Stat. &\\
        \midrule
        GPT-4
        & 0.0 & 0.5 & -0.5 & 1.0 &  \\
        
        Claude 3
        & 20.0 & 20.0 & 0.0 & / &  \\

        LLaMA 2
        & 14.5 & 8.0 & 6.5 & 4.0 & \starx \\
        
        Mixtral
        & 0.0 & 1.8 & -1.8 & 0.2 & \\

                \toprule
         \multicolumn{6}{l}{\textbf{Story 2. Supermarket:}  Sign \textit{vs.} Not sign} \\
         \midrule
         & \multicolumn{2}{c|}{FCE per group} & \multicolumn{3}{c}{Kruskal-Wallis}\\
         \cmidrule{2-3}\cmidrule{4-6}
         & Male & Female & Diff. & Stat. &\\
        \midrule
        GPT-4
        & 5.5 & 11.0 & -5.5 & 3.0 &  \\
        
        Claude 3
        & 26.0 & 25.0 & 1.0 & 0.0 &  \\

        LLaMA 2
        & 0.5 & 1.5 & -1.0 & 1.1 &  \\
        
        Mixtral
        & 39.0 & 42.0 & -3.0 & 0.3 & \\

        \toprule
         \multicolumn{6}{l}{\textbf{Story 3. Traffic Ticket:}  Pay fine \textit{vs.} Contest} \\
         \midrule
         & \multicolumn{2}{c|}{FCE per group} & \multicolumn{3}{c}{Kruskal-Wallis}\\
         \cmidrule{2-3}\cmidrule{4-6}
         & Male & Female & Diff. & Stat. &\\
        \midrule
        GPT-4
        & 5.0 & 0.0 &  5.0 & 13.0  & \starxxx  \\
        
        Claude 3
        & 0.0 & 0.0 & 0.0 & / &  \\

        LLaMA 2
        & 0.0 & 0.0 & 0.0 & / &  \\
        
        Mixtral
        & 11.0 & 7.5 & 3.5 & 1.4 & \\

        \toprule
         \multicolumn{6}{l}{\textbf{Story 4. Space R\&D:}  Vote for \textit{vs.} Vote against} \\
         \midrule
         & \multicolumn{2}{c|}{FCE per group} & \multicolumn{3}{c}{Kruskal-Wallis}\\
         \cmidrule{2-3}\cmidrule{4-6}
         & Male & Female & Diff. & Stat. &\\
        \midrule
        GPT-4
        & 20.0 & 20.0 &  0.0 & / &  \\
        
        Claude 3
        & 8.0 & 10.3 & -2.3 & 0.6 &  \\

        LLaMA 2
        & 17.5 & 18.0 & -0.5 & 0.0 &  \\
        
        Mixtral
        & 2.3 & 4.3 & -2.0 & 0.0 & \\

        \toprule
        \multicolumn{6}{r}{{\small \starx $p < 0.05$, \starxx $p < 0.01$, \starxxx $p < 0.001$}}
    \end{tabular}
    \caption{Kruskal-Wallis test for H1-3 (Gender). \textit{Diff.} and \textit{Stat.} columns indicates agreement difference between two options and $H$-statistics from the test, respectively.}
    \label{tab:h1-3-gender}
\end{table}

%% file: table/h2.tex
\begin{table*}[tp]
    \centering
    \small
    \begin{tabular}{lr@{\ \;}r@{\ \;}r@{\ \;}rr@{}l@{\ \;}p{8em}r@{\ \;}r@{\ \;}r@{\ \;}rr@{}l@{\ \;}p{8em}}
        \toprule
         \multicolumn{15}{l}{\textbf{H2-1.} The FCE of supportive information is higher than the others.}\\
        \midrule
         & \multicolumn{7}{l}{\textbf{Story 2. Supermarket}: Sign \textit{vs.} Not sign} & \multicolumn{7}{l}{\textbf{Story 4. Space R\&D}: Vote for \textit{vs.} Vote against} \\
         & P1 & P2 & P3 & P4 & \multicolumn{2}{c}{K-W test} & {M-W tests}
         & P1 & P2 & P3 & P4 & \multicolumn{2}{c}{K-W test} & {M-W tests} \\
        \cmidrule(r){2-8}\cmidrule(r){9-15}
        \gpt    
        & 8.3 & 20.0 & -9.5 & 19.5 & 120.9 & \starxxx & \texttt{\small P2>P1>P3, P4>P1}
        & 20.0 & 20.0 & 0.0 & 20.0 & 159.0 & \starxxx & \texttt{\small P2>P3,P1>P3,P4>P3} \\
        
        \claude    
        & 25.5 & 29.0 & -29.5 & 27.7 & 97.6 & \starxxx & \texttt{\small P2>P3,P1>P3,P4>P3}
        & 9.1 & 5.0 & 0.0 & 20.4 & 63.9 & \starxxx & \texttt{\small P1>P3>P4, P4>P2} \\

        \llama    
        & 1.0 & 20.0 & -3.3 & 37.0 & 125.1 & \starxxx & \texttt{\small P2>P1>P3,P4>P2>P3}
        & 17.8 & 20.0 & -19.0 & 12.5 & 120.1 & \starxxx & \texttt{\small P2>P3>P4, P1>P3} \\
        
        \mixtral    
        & 40.5 & 47.9 & -33.0 & 40.0 & 97.3 & \starxxx & \texttt{\small P2>P3, P1>P3}
        & 3.3 & 0.0 & 1.6 & 0.5 & 103.9 & \starxxx & \texttt{\small P1>P3>P4, P1>P2} \\

        \toprule
         \multicolumn{15}{l}{\textbf{H2-2.} Deeper reasoning decreases FCE.}\\
         \midrule
         & \multicolumn{7}{l}{\textbf{Story 2. Supermarket}: Sign \textit{vs.} Not sign} & \multicolumn{7}{l}{\textbf{Story 4. Space R\&D}: Vote for \textit{vs.} Vote against} \\
         & R1 & R2 & R3 & R4 & \multicolumn{2}{c}{K-W test} & {M-W tests}
         & R1 & R2 & R3 & R4 & \multicolumn{2}{c}{K-W test} & {M-W tests} \\
        \cmidrule(r){2-8}\cmidrule(r){9-15}

        \gpt    
        & 8.3 & 9.5 & 8.0 & 24.5 & 26.2 & \starxxx & \texttt{\small R4>R1,R4>R2,R4>R3}
        & 20.0 & 20.0 & 0.3 & 7.5 & 159.0 & \starxxx & \texttt{\small R1>R4,R2>R4,R4>R3} \\
        
        \claude    
        & 25.2 & 25.5 & 6.6 & 2.3 & 103.0 & \starxxx & \texttt{\small R1>R3,R1>R4,R2>R3, R2>R4}
        & 9.1 & 20.3 & 15.9 & -0.9 & 96.5 & \starxxx & \texttt{\small R3>R2>R1>R4} \\

        \llama    
        & 1.0 & -0.9 & 2.8 & 6.5 & 3.8 &  & \texttt{\small -}
        & 17.8 & 11.5 & 2.1 & 5.8 & 49.4 & \starxxx & \texttt{\small R1>R3, R1>R2>R4} \\
        
        \mixtral   
        & 40.5 & 32.6 & 14.6 & 46.6 & 73.0 & \starxxx & \texttt{\small R1>R2>R3,R2>R3>R4}
        & 3.3 & 5.1 & 2.7 & 5.4 & 1.3 &  & \texttt{\small -} \\
        
        \bottomrule
         \multicolumn{15}{r}{ \starx $p < 0.05$, \starxx $p < 0.01$, \starxxx $p < 0.001$}\\
    \end{tabular}
    \caption{Kruskal-Wallis Test for H2. Rows \gpt, \claude, \llama, \mixtral\ indicates GPT-4, Claude 3, LLaMA 2, and Mixtral. K-W test and M-W test indicates Kruskal-Wallis test statistics and pairwise Mann-Whitney test results, respectively.}
    \label{tab:h2-summary}
\end{table*}

%% file: table/normality.tex
\begin{table*}[tp]
    \centering
    \small
    \begin{tabular}{cclrlrclrlr}
      \toprule
      Model & Story & \multicolumn{2}{l}{Option 1} & \multicolumn{2}{l}{Option 2} & Story & \multicolumn{2}{l}{Option 1} & \multicolumn{2}{l}{Option 2} \\
      \cmidrule(lr){3-4}\cmidrule(lr){5-6}\cmidrule(lr){8-9}\cmidrule(lr){10-11}
      && $W$ & $p$ & $W$ & $p$ && $W$ & $p$ & $W$ & $p$ \\
      \cmidrule(lr){1-6}\cmidrule(lr){7-11}
      GPT-4 & Story 1 & 1.0   &    1.000 & 0.147 & $<$0.001
       & Story 2 & 0.147 & $<$0.001 & 0.623 & $<$0.001 \\
       & Story 3 & 0.539 & $<$0.001 & 1.0   &    1.000
       & Story 4 & 1.0   &    1.000 & 1.0   &    1.000 \\
      \midrule
      Claude 3 & Story 1 & 1.0   &    1.000 & 1.0   &    1.000 
       & Story 2 & 0.345 & $<$0.001 & 0.634 & $<$0.001 \\
       & Story 3 & 1.0   &    1.000 & 1.0   &    1.000 
       & Story 4 & 1.0   &    1.000 & 0.66  & $<$0.001 \\
      \midrule
      LLaMA 2 & Story 1 & 0.147 & $<$0.001 & 0.634 & $<$0.001 
       & Story 2 & 1.0   &    1.000 & 0.345 & $<$0.001 \\
       & Story 3 & 1.0   &    1.000 & 1.0   &    1.000 
       & Story 4 & 0.389 & $<$0.001 & 0.462 & $<$0.001 \\
      \midrule
      Mixtral & Story 1 & 0.845 & $<$0.001 & 0.335 & $<$0.001 
       & Story 2 & 0.631 & $<$0.001 & 0.772 & $<$0.001 \\
       & Story 3 & 0.582 & $<$0.001 & 0.772 & $<$0.001 
       & Story 4 & 0.674 & $<$0.001 & 0.78  & $<$0.001 \\
      \bottomrule
    \end{tabular}
    \caption{Shapiro-Wilk normality test result on our experimental result}
    \label{tab:normality}
\end{table*}

%% file: table/human-compare.tex
\begin{table*}
    \centering
    \begin{tabular}{@{}lc@{\;}c@{\;}r@{}lc@{\;}c@{\;}r@{}lc@{\;}c@{\;}r@{}lc@{\;}c@{\;}r@{}l@{}}
        \toprule
         & \multicolumn{4}{l}{\small \textbf{Story 1. Term paper}} & \multicolumn{4}{l}{\small \textbf{Story 2. Supermarket}} & \multicolumn{4}{l}{\small \textbf{Story 3. Traffic Ticket}} & \multicolumn{4}{l}{\small \textbf{Story 4. Space R\&D}} \\
         & \multicolumn{4}{l}{\small Individual \textit{vs}. Group} & \multicolumn{4}{l}{\small Sign \textit{vs}. Not sign} & \multicolumn{4}{l}{\small Pay fine \textit{vs}. Contest} & \multicolumn{4}{l}{\small Vote for \textit{vs}. Vote against} \\
        \cmidrule(r){2-5}\cmidrule(r){6-9}\cmidrule(r){10-13}\cmidrule(r){14-17}
         & $\mu_1$ & $\mu_2$ & \multicolumn{2}{c}{Diff.} & $\mu_1$ & $\mu_2$ & \multicolumn{2}{c}{Diff.} & $\mu_1$ & $\mu_2$ & \multicolumn{2}{c}{Diff.} & $\mu_1$ & $\mu_2$ & \multicolumn{2}{c}{Diff.}\\
        
        \midrule
        \multicolumn{17}{l}{\textbf{LLMs}: Our experimental result of FCE}\\
        \midrule
        GPT-4
        & 60.0 & 59.7 &  0.3 &  
        & 60.3 & 52.0 &  8.3 & \starxxx 
        & 62.5 & 60.0 &  2.5 & \starxxx 
        & 60.0 & 40.0 & 20.0 & \starxxx \\
        
        Claude 3
        & 60.0 & 40.0 & 20.0 & \starxxx 
        & 61.0 & 35.5 & 25.5 & \starxxx 
        & 70.0 & 70.0 &  0.0 &  
        & 60.0 & 50.9 &  9.1 & \starxxx \\

        LLaMA 2
        & 60.3 & 49.0 & 11.3 & \starxxx 
        & 70.0 & 69.0 &  1.0 & \starx 
        & 70.0 & 70.0 &  0.0 &  
        & 61.3 & 43.5 & 17.8 & \starxxx \\
        
        Mixtral
        & 60.9 & 60.0 &  0.9 & 
        & 76.3 & 35.8 & 40.5 & \starxxx 
        & 65.8 & 56.5 &  9.3 & \starxxx 
        & 42.9 & 39.6 &  3.3 & \starxxx \\

        \midrule
        \multicolumn{17}{l}{\textbf{Human}: Result of American students, provided by \citet{ross1977false}}\\
        \midrule
        American
        & 67.4 & 45.9 & 21.5 & \starxxx 
        & 75.6 & 57.3 & 18.3 & \starxxx 
        & 71.8 & 51.7 & 20.1 & \starxxx 
        & 65.7 & 48.5 & 17.2 & \starx \\ 

        \midrule
        \multicolumn{17}{l}{\textbf{Human}: Result of American/Korean students for Story 1 and 2, provided by \citet{cross-culture}}\\
        \midrule
        Korean
        & 67.10 & 38.33 & 28.77 & \starxxx
        & 71.46 & 32.20 & 39.26 & \starxxx
        & - & - & - & 
        & - & - & - & \\

        American
        & 69.21 & 55.06 & 14.15 & \starxx
        & 78.59 & 69.72 & 8.87 & \starx
        & - & - & - & 
        & - & - & - & \\
        \bottomrule
        \multicolumn{17}{r}{{\small \starx $p < 0.05$, \starxx $p < 0.01$, \starxxx $p < 0.001$}}
    \end{tabular}
    \caption{Comparing our experimental result on LLMs with previous reports on humans. \textit{Diff.} columns indicate the mean value of $\mu_1-\mu_2$ and the $p$-values of the Mann-Whitney test.}
    \label{tab:human}
\end{table*}

%% file: table/freeform.tex
\begin{table}[!t]
    \centering
    \small
    \begin{tabular}{@{}lcc|cc@{}l@{}}
        \toprule
         \multicolumn{6}{l}{\textbf{Story 1. Term paper:}  Individual \textit{vs.} Group} \\
         \midrule
         & \multicolumn{2}{c|}{Perceived Agreement on} & \multicolumn{3}{c}{Mann-Whitney}\\
         \cmidrule{2-3}\cmidrule{4-6}
         & Individual & Group & Diff. & Stat. &\\
        \midrule
        Mixtral
        & 60.3 & 60.0 &  0.3 & 57 & \\

        \toprule
         \multicolumn{6}{l}{\textbf{Story 3. Traffic Ticket:}  Pay fine \textit{vs.} Contest} \\
         \midrule
         & \multicolumn{2}{c|}{Perceived Agreement on} & \multicolumn{3}{c}{Mann-Whitney}\\
         \cmidrule{2-3}\cmidrule{4-6}
         & Pay fine & Contest & Diff. & Stat. &\\
        \midrule

        Claude 3
        & 70.0 & 69.0 & 1.0 & 220 &  \\

        \toprule
         \multicolumn{6}{l}{\textbf{Story 4. Space R\&D:}  Vote for \textit{vs.} Vote against} \\
         \midrule
         & \multicolumn{2}{c|}{Perceived Agreement on} & \multicolumn{3}{c}{Mann-Whitney}\\
         \cmidrule{2-3}\cmidrule{4-6}
         & Vote for & Vote against & Diff. & Stat. &\\
        \midrule
        GPT-4
        & 60.0 & 40.0 & 20.0 & 263 & \starxxx \\

        LLaMA 2
        & 60.5 & 40.0 & 20.5 & 279 & \starxxx \\

        \toprule
        \multicolumn{6}{r}{{\small \starx $p < 0.05$, \starxx $p < 0.01$, \starxxx $p < 0.001$}}
    \end{tabular}
    \caption{Mann-Whitney U Test for free form. \textit{Diff.} and \textit{Stat.} columns indicates the agreement difference between two options and $U$-statistics from the test, respectively. Human results are copied from \citep{ross1977false}.}
    \label{tab:free-form}
\end{table}

%% file: table/h1-america.tex
\begin{table}[!t]
    \centering
    \small
    \begin{tabular}{lcc|rr@{}l}
        \toprule
         \multicolumn{6}{l}{\textbf{Story 1. Term paper:}  Individual \textit{vs.} Group} \\
         \midrule
         & \multicolumn{2}{c|}{Perceived Agreement on} & \multicolumn{3}{c}{Mann-Whitney}\\
         \cmidrule{2-3}\cmidrule{4-6}
         & Individual & Group & Diff. & Stat. &\\
        \midrule
        GPT-4
        & 60.0 & 59.5 &  0.5 & 210 &  \\
        
        Claude 3
        & 60.0 & 40.0 & 20.0 & 400 & \starxxx \\

        LLaMA 2
        & 60.0 & 55.0 & 5.0 & 250 & \starxx \\
        
        Mixtral
        & 59.8 & 59.0 & 0.8 & 210 & \\

                \toprule
         \multicolumn{6}{l}{\textbf{Story 2. Supermarket:}  Sign \textit{vs.} Not sign} \\
         \midrule
         & \multicolumn{2}{c|}{Perceived Agreement on} & \multicolumn{3}{c}{Mann-Whitney}\\
         \cmidrule{2-3}\cmidrule{4-6}
         & Sign & Not sign & Diff. & Stat. &\\
        \midrule
        GPT-4
        & 60.0 & 59.0 & 1.0 & 210 &  \\
        
        Claude 3
        & 60.0 & 40.0 & 20.0 & 400 & \starxxx \\

        LLaMA 2
        & 70.0 & 68.0 & 2.0 & 240 & \starx \\
        
        Mixtral
        & 73.5 & 33.5 & 40.0 & 393 & \starxxx\\

        \toprule
         \multicolumn{6}{l}{\textbf{Story 3. Traffic Ticket:}  Pay fine \textit{vs.} Contest} \\
         \midrule
         & \multicolumn{2}{c|}{Perceived Agreement on} & \multicolumn{3}{c}{Mann-Whitney}\\
         \cmidrule{2-3}\cmidrule{4-6}
         & Pay fine & Contest & Diff. & Stat. &\\
        \midrule
        GPT-4
        & 63.0 & 60.0 &  3.0 & 260 & \starxx  \\
        
        Claude 3
        & 70.0 & 70.0 & 0.0 & 200 &  \\

        LLaMA 2
        & 70.0 & 70.0 & 0.0 & 200 &  \\
        
        Mixtral
        & 62.0 & 52.0 & 10.0 & 302 & \starxxx \\

        \toprule
         \multicolumn{6}{l}{\textbf{Story 4. Space R\&D:}  Vote for \textit{vs.} Vote against} \\
         \midrule
         & \multicolumn{2}{c|}{Perceived Agreement on} & \multicolumn{3}{c}{Mann-Whitney}\\
         \cmidrule{2-3}\cmidrule{4-6}
         & Vote & Vote against & Diff. & Stat. &\\
        \midrule
        GPT-4
        & 60.0 & 40.0 & 20.0 & 400 & \starxxx \\
        
        Claude 3
        & 60.0 & 42.8 & 17.2 & 370 & \starxxx \\

        LLaMA 2
        & 60.0 & 41.0 & 19.0 & 390 & \starxxx \\
        
        Mixtral
        & 43.8 & 39.3 & 4.5 & 301 & \starxx \\

        \toprule
        \multicolumn{6}{r}{{\small \starx $p < 0.05$, \starxx $p < 0.01$, \starxxx $p < 0.001$}}
    \end{tabular}
    \caption{Result of Mann-Whitney U Test, on \textbf{European American Group}.}
    \label{tab:h1-american}
\end{table}

%% file: table/h1-korea.tex
\begin{table}[!t]
    \centering
    \small
    \begin{tabular}{lcc|rr@{}l}
        \toprule
         \multicolumn{6}{l}{\textbf{Story 1. Term paper:}  Individual \textit{vs.} Group} \\
         \midrule
         & \multicolumn{2}{c|}{Perceived Agreement on} & \multicolumn{3}{c}{Mann-Whitney}\\
         \cmidrule{2-3}\cmidrule{4-6}
         & Individual & Group & Diff. & Stat. &\\
        \midrule
        GPT-4
        & 60.0 & 60.0 &  0.0 & 200 &  \\
        
        Claude 3
        & 60.0 & 40.0 & 20.0 & 400 & \starxxx \\

        LLaMA 2
        & 60.0 & 43.0 & 17.5 & 371 & \starxxx \\
        
        Mixtral
        & 62.0 & 61.0 & 1.0 & 239 & \\

                \toprule
         \multicolumn{6}{l}{\textbf{Story 2. Supermarket:}  Sign \textit{vs.} Not sign} \\
         \midrule
         & \multicolumn{2}{c|}{Perceived Agreement on} & \multicolumn{3}{c}{Mann-Whitney}\\
         \cmidrule{2-3}\cmidrule{4-6}
         & Sign & Not sign & Diff. & Stat. &\\
        \midrule
        GPT-4
        & 60.5 & 45.0 & 15.5 & 352 & \starxxx \\
        
        Claude 3
        & 62.0 & 31.0 & 31.0 & 400 & \starxxx \\

        LLaMA 2
        & 70.0 & 70.0 & 0.0 & 200 &  \\
        
        Mixtral
        & 79.0 & 38.0 & 41.0 & 398 & \starxxx\\

        \toprule
         \multicolumn{6}{l}{\textbf{Story 3. Traffic Ticket:}  Pay fine \textit{vs.} Contest} \\
         \midrule
         & \multicolumn{2}{c|}{Perceived Agreement on} & \multicolumn{3}{c}{Mann-Whitney}\\
         \cmidrule{2-3}\cmidrule{4-6}
         & Pay fine & Contest & Diff. & Stat. &\\
        \midrule
        GPT-4
        & 62.0 & 60.0 &  2.0 & 240 & \starx  \\
        
        Claude 3
        & 70.0 & 70.0 & 0.0 & 200 &  \\

        LLaMA 2
        & 70.0 & 70.0 & 0.0 & 200 &  \\
        
        Mixtral
        & 69.5 & 61.0 & 8.5 & 331 & \starxxx \\

        \toprule
         \multicolumn{6}{l}{\textbf{Story 4. Space R\&D:}  Vote for \textit{vs.} Vote against} \\
         \midrule
         & \multicolumn{2}{c|}{Perceived Agreement on} & \multicolumn{3}{c}{Mann-Whitney}\\
         \cmidrule{2-3}\cmidrule{4-6}
         & Vote & Vote against & Diff. & Stat. &\\
        \midrule
        GPT-4
        & 60.0 & 40.0 & 20.0 & 400 & \starxxx \\
        
        Claude 3
        & 60.0 & 59.0 & 1.0 & 210 &  \\

        LLaMA 2
        & 62.5 & 46.0 & 16.5 & 355 & \starxxx \\
        
        Mixtral
        & 42.0 & 40.0 & 2.0 & 245 &  \\

        \toprule
        \multicolumn{6}{r}{{\small \starx $p < 0.05$, \starxx $p < 0.01$, \starxxx $p < 0.001$}}
    \end{tabular}
    \caption{Result of Mann-Whitney U Test, on \textbf{Korean Group}.}
    \label{tab:h1-korean}
\end{table}

%% file: table/h1-man.tex
\begin{table}[!t]
    \centering
    \small
    \begin{tabular}{lcc|rr@{}l}
        \toprule
         \multicolumn{6}{l}{\textbf{Story 1. Term paper:}  Individual \textit{vs.} Group} \\
         \midrule
         & \multicolumn{2}{c|}{Perceived Agreement on} & \multicolumn{3}{c}{Mann-Whitney}\\
         \cmidrule{2-3}\cmidrule{4-6}
         & Individual & Group & Diff. & Stat. &\\
        \midrule
        GPT-4
        & 60.0 & 60.0 &  0.0 & 200 &  \\
        
        Claude 3
        & 60.0 & 40.0 & 20.0 & 400 & \starxxx \\

        LLaMA 2
        & 60.5 & 46.0 & 14.5 & 343 & \starxxx \\
        
        Mixtral
        & 61.0 & 61.0 &  0.9 & 222 & \\

                \toprule
         \multicolumn{6}{l}{\textbf{Story 2. Supermarket:}  Sign \textit{vs.} Not sign} \\
         \midrule
         & \multicolumn{2}{c|}{Perceived Agreement on} & \multicolumn{3}{c}{Mann-Whitney}\\
         \cmidrule{2-3}\cmidrule{4-6}
         & Sign & Not sign & Diff. & Stat. &\\
        \midrule
        GPT-4
        & 60.5 & 55.0 & 5.5 & 257 & \starxx \\
        
        Claude 3
        & 62.0 & 36.0 & 26.0 & 400 & \starxxx \\

        LLaMA 2
        & 70.0 & 69.5 & 0.5 & 210 &  \\
        
        Mixtral
        & 77.0 & 38.0 & 39.0 & 394 & \starxxx\\

        \toprule
         \multicolumn{6}{l}{\textbf{Story 3. Traffic Ticket:}  Pay fine \textit{vs.} Contest} \\
         \midrule
         & \multicolumn{2}{c|}{Perceived Agreement on} & \multicolumn{3}{c}{Mann-Whitney}\\
         \cmidrule{2-3}\cmidrule{4-6}
         & Pay fine & Contest & Diff. & Stat. &\\
        \midrule
        GPT-4
        & 65.0 & 60.0 &  5.0 & 300 & \starxxx \\
        
        Claude 3
        & 70.0 & 70.0 & 0.0 & 200 &  \\

        LLaMA 2
        & 70.0 & 70.0 & 0.0 & 200 &  \\
        
        Mixtral
        & 68.0 & 57.0 & 11.0 & 334 & \starxxx \\

        \toprule
         \multicolumn{6}{l}{\textbf{Story 4. Space R\&D:}  Vote for \textit{vs.} Vote against} \\
         \midrule
         & \multicolumn{2}{c|}{Perceived Agreement on} & \multicolumn{3}{c}{Mann-Whitney}\\
         \cmidrule{2-3}\cmidrule{4-6}
         & Vote & Vote against & Diff. & Stat. &\\
        \midrule
        GPT-4
        & 60.0 & 40.0 & 20.0 & 400 & \starxxx \\
        
        Claude 3
        & 60.0 & 52.0 & 8.0 & 280 & \starxxx \\

        LLaMA 2
        & 62.5 & 45.0 & 17.5 & 363 & \starxxx \\
        
        Mixtral
        & 42.3 & 40.0 & 2.3 & 268 & \starx \\

        \bottomrule
        \multicolumn{6}{r}{{\small \starx $p < 0.05$, \starxx $p < 0.01$, \starxxx $p < 0.001$}}
    \end{tabular}
    \caption{Mann-Whitney U Test for \textbf{Male Group}.}
    \label{tab:h1-male}
\end{table}

%% file: table/h1-woman.tex
\begin{table}[!t]
    \centering
    \small
    \begin{tabular}{lcc|rr@{}l}
        \toprule
         \multicolumn{6}{l}{\textbf{Story 1. Term paper:}  Individual \textit{vs.} Group} \\
         \midrule
         & \multicolumn{2}{c|}{Perceived Agreement on} & \multicolumn{3}{c}{Mann-Whitney}\\
         \cmidrule{2-3}\cmidrule{4-6}
         & Individual & Group & Diff. & Stat. &\\
        \midrule
        GPT-4
        & 60.0 & 59.5 &  0.5 & 200 &  \\
        
        Claude 3
        & 60.0 & 40.0 & 20.0 & 400 & \starxxx \\

        LLaMA 2
        & 60.0 & 52.0 & 8.0 & 280 & \starxxx \\
        
        Mixtral
        & 60.8 & 59.0 & 1.8 & 229 & \\

                \toprule
         \multicolumn{6}{l}{\textbf{Story 2. Supermarket:}  Sign \textit{vs.} Not sign} \\
         \midrule
         & \multicolumn{2}{c|}{Perceived Agreement on} & \multicolumn{3}{c}{Mann-Whitney}\\
         \cmidrule{2-3}\cmidrule{4-6}
         & Sign & Not sign & Diff. & Stat. &\\
        \midrule
        GPT-4
        & 60.0 & 49.0 & 11.0 & 310 & \starxxx \\
        
        Claude 3
        & 60.0 & 35.0 & 25.0 & 400 & \starxxx \\

        LLaMA 2
        & 70.0 & 68.5 & 1.5 & 230 & \starx \\
        
        Mixtral
        & 75.5 & 33.5 & 42.0 & 395 & \starxxx\\

        \toprule
         \multicolumn{6}{l}{\textbf{Story 3. Traffic Ticket:}  Pay fine \textit{vs.} Contest} \\
         \midrule
         & \multicolumn{2}{c|}{Perceived Agreement on} & \multicolumn{3}{c}{Mann-Whitney}\\
         \cmidrule{2-3}\cmidrule{4-6}
         & Pay fine & Contest & Diff. & Stat. &\\
        \midrule
        GPT-4
        & 60.0 & 60.0 &  0.0 & 200 &  \\
        
        Claude 3
        & 70.0 & 70.0 & 0.0 & 200 &  \\

        LLaMA 2
        & 70.0 & 70.0 & 0.0 & 200 &  \\
        
        Mixtral
        & 63.5 & 56.0 & 7.5 & 282 & \starxx \\

        \toprule
         \multicolumn{6}{l}{\textbf{Story 4. Space R\&D:}  Vote for \textit{vs.} Vote against} \\
         \midrule
         & \multicolumn{2}{c|}{Perceived Agreement on} & \multicolumn{3}{c}{Mann-Whitney}\\
         \cmidrule{2-3}\cmidrule{4-6}
         & Vote & Vote against & Diff. & Stat. &\\
        \midrule
        GPT-4
        & 60.0 & 40.0 & 20.0 & 400 & \starxxx \\
        
        Claude 3
        & 60.0 & 49.8 & 10.2 & 300 & \starxxx \\

        LLaMA 2
        & 60.0 & 42.0 & 18.0 & 380 & \starxxx \\
        
        Mixtral
        & 43.5 & 39.3 & 4.3 & 278 & \starx \\

        \bottomrule
        \multicolumn{6}{r}{{\small \starx $p < 0.05$, \starxx $p < 0.01$, \starxxx $p < 0.001$}}
    \end{tabular}
    \caption{Mann-Whitney U Test for \textbf{Female Group}.}
    \label{tab:h1-female}
\end{table}

%% file: table/range.tex
\begin{table*}[t]
    \centering
    \begin{tabular}{l|rrrrrrr}
     \toprule
       & \multicolumn{7}{c}{Perceived agreement value answered by LLMs} \\
     \cmidrule(lr){2-8}
       & $<$10 & 10-20 & 20-30 
       & 30-70 & 70-80 & 80-90 & $\ge$ 90 \\
     \midrule
     GPT-4
       & 0 & 0 & 4 & 5,030 & 86 & 0 & 0 \\

       & & \multicolumn{1}{c|}{} & & {\small (98.2\%)} & \multicolumn{1}{c|}{} & & \\
     \cmidrule{4-6}
       & & & & {\small (100.0\%)} & & & \\

     Claude 3
       & 0 & 5 & 73 & 4,731 & 130 & 0 & 0 \\

       & \multicolumn{1}{c|}{} & \multicolumn{1}{c|}{} & & {\small (92.4\%)} & \multicolumn{1}{c|}{} & \multicolumn{1}{c|}{} & \\
     \cmidrule{4-6}
       & \multicolumn{1}{c|}{}& & \multicolumn{3}{c}{\small (96.4\%)} & \multicolumn{1}{c|}{} & \\
     \cmidrule{3-7}
       & & \multicolumn{5}{c}{\small (100.0\%)} & \\
       
     LLaMA 2
       & 1 & 42 & 104 & 4,527 & 169 & 38 & 0\\

       & \multicolumn{1}{c|}{} & \multicolumn{1}{c|}{} & & {\small (92.7\%)} & \multicolumn{1}{c|}{} & \multicolumn{1}{c|}{} & \\
     \cmidrule{4-6}
       & \multicolumn{1}{c|}{}& & \multicolumn{3}{c}{\small (98.3\%)} & \multicolumn{1}{c|}{} & \\
     \cmidrule{3-7}
       & & \multicolumn{5}{c}{\small (99.9\%)} & \\
       
     Mixtral
       & 0 & 7 & 183 & 4,280 & 596 & 54 & 0 \\
       
       & \multicolumn{1}{c|}{} & \multicolumn{1}{c|}{} & & {\small (83.6\%)} & \multicolumn{1}{c|}{} & \multicolumn{1}{c|}{} & \\
     \cmidrule{4-6}
       & \multicolumn{1}{c|}{}& & \multicolumn{3}{c}{\small (98.8\%)} & \multicolumn{1}{c|}{} & \\
     \cmidrule{3-7}
       & & \multicolumn{5}{c}{\small (100.0\%)} & \\
    \bottomrule
    
    \end{tabular}
    \caption{The percentage of answers in a particular range. Rows \gpt, \claude, \llama, \mixtral\ indicates GPT-4, Claude 3, LLaMA 2, and Mixtral.}
    \label{tab:h2-range}
\end{table*}

%% file: table/h2_appendix.tex
\begin{table*}[tp]
    \centering
    \small
    \begin{tabular}{lr@{\ \;}r@{\ \;}r@{\ \;}rr@{}l@{\ \;}p{8em}r@{\ \;}r@{\ \;}r@{\ \;}rr@{}l@{\ \;}p{8em}}
        \toprule
         \multicolumn{15}{l}{\textbf{H2-1.} The FCE of supportive information is higher than the others.}\\
        \midrule
         & \multicolumn{7}{l}{\textbf{Story 1. Term paper}: Individual \textit{vs.} Group} & \multicolumn{7}{l}{\textbf{Story 3. Traffic Ticket}: Pay fine \textit{vs.} Contest} \\
         & P1 & P2 & P3 & P4 & $H$ & & {\small Mann-Whitney test}
         & P1 & P2 & P3 & P4 & $H$ & & {\small Mann-Whitney test} \\
        \cmidrule(r){2-8}\cmidrule(r){9-15}
        \gpt    
        & 0.3 & 0.0 & 0.0 & -1.0 & 11.9 & \starxx & \texttt{\small P2 > P4}
        & 2.5 & 10.0 & -9.5 & 10.0 & 141.4 & \starxxx & \texttt{\small P2 > P1, P3} \\
        
        \claude    
        & 20.0 & 21.3 & 0.0 & 0.0 & 154.8 & \starxxx & \texttt{\small P2 > P1, P3, P4}
        & 0.0 & 0.0 & 0.0 & 0.0 & - &  & {\small      -} \\

        \llama    
        & 11.3 & 0.0 & 0.0 & 3.4 & 45.9 & \starxxx & {\small -}
        & 0.0 & 10.0 & 0.0 & 0.0 & 159 & \starxxx & \texttt{\small P2 > P1, P3, P4} \\
        
        \mixtral    
        & 0.9 & 7.0 & 27.3 & 0.3 & 99.9 & \starxxx & \texttt{\small P2 > P1, P3}
        & 9.3 & 10.5 & -31.0 & 9.0 & 103.9 & \starxxx & \texttt{\small P2 > P1, P3, P4} \\

        \toprule
         \multicolumn{15}{l}{\textbf{H2-2.} Deeper reasoning decreases FCE.}\\
         \midrule
         & \multicolumn{7}{l}{\textbf{Story 2. Supermarket}: Sign \textit{vs.} Not sign} & \multicolumn{7}{l}{\textbf{Story 4. Space R\&D}: Vote for \textit{vs.} Vote against} \\
         & R1 & R2 & R3 & R4 & $H$ & & {\small Mann-Whitney test}
         & R1 & R2 & R3 & R4 & $H$ & & {\small Mann-Whitney test} \\
        \cmidrule(r){2-8}\cmidrule(r){9-15}
        \gpt    
        & 0.3 & 0.0 & -20.0 & 0.0 & 156.1 & \starxxx & \texttt{\small R1 > R2 > R3}
        & 2.5 & 10.0 & 0.0 & -2.8 & 79.3 & \starxxx & \texttt{\small R2 > R3 > R4} \\
        
        \claude    
        & 20.0 & 13.5 & 1.3 & -40.0 & 137.9 & \starxxx & \texttt{\small R1 > R2 > R3 > R4}
        & 0.0 & 0.0 & 4.9 & -48.8 & 132.9 & \starxxx & \texttt{\small R3 > R4} \\

        \llama
        & 11.3 & 7.9 & 4.9 & -0.7 & 11.9 & \starxx & \texttt{\small R1 > R2 > R3 > R4}
        & 0.0 & 0.1 & -8.4 & 21.1 & 51.5 & \starxxx & \texttt{\small R2 > R3} \\
        
        \mixtral
        & 0.9 & 7.3 & 6.6 & 5.9 & 6.9 &  &\texttt{\small R2 > R3 > R4}
        & 9.3 & 6.0 & 5.9 & 19.9 & 42.3 & \starxxx & \texttt{\small R1 > R2 > R3} \\
        
        \toprule
        \multicolumn{15}{r}{\starx $p<0.05$ \starxx $p<0.01$ \starxxx $p<0.001$}
    \end{tabular}
    \caption{Kruskal-Wallis Test for H2. Rows \gpt, \claude, \llama, \mixtral\ indicates GPT-4, Claude 3, LLaMA 2, and Mixtral. $H$ column indicates Kruskal-Wallis test statistics.}
    \label{tab:h2-summary-app}
\end{table*}

%% file: table/story1.tex
\begin{table*}
    \centering
    \begin{tabular}{@{}lcrcrrcr@{}lcr@{}l@{}}
        \toprule
         \multicolumn{12}{l}{\textbf{Story 1. Term paper}}\\
        \toprule
         & \multicolumn{2}{l}{Condition 1} & \multicolumn{2}{l}{Condition 2} & Diff. & \multicolumn{3}{l}{Dunn's post-test} & \multicolumn{3}{l}{Mann-Whitney test} \\
         \cmidrule(r){2-3}\cmidrule(r){4-5}\cmidrule(r){7-9}\cmidrule(r){10-12}
         &  & FCE &  & FCE &  & Hypothesis & $p$-value & & Hypothesis & $U$ & \\
         \cmidrule(r){2-12}
        \gpt\ GPT-4 
         & P1 & 0.3 & P2 & 0.0 & +0.3 & P1 $\ne$ P2? & 0.532 &  & - &  &  \\
         &    &     & P3 & 0.0 & +0.3 & P1 $\ne$ P3? & 0.532 &  & - &  &  \\
         &    &     & P4 & -1.0 & +1.3 & P1 $\ne$ P4? & 0.002 & \starxx & P1 $>$ P4 & 898 & \starx \\
         & P2 & 0.0 & P3 & 0.0 & 0.0 & P2 $\ne$ P3? & 1.000 &  & - &  &  \\
         &    &     & P4 & -1.0 & +1.0 & P2 $\ne$ P4? & 0.011 & \starx & P2 $>$ P4 & 880 & \starx \\
         & P3 & 0.0 & P4 & -1.0 & +1.0 & P3 $\ne$ P4? & 0.011 & \starx & P3 $>$ P4 & 880 & \starx \\
        \cmidrule(r){2-12}
        \claude\ Claude 3
         & P1 & 20.0 & P2 & 21.3 & -1.3 & P1 $\ne$ P2? & 0.583 &  & - &  &  \\
         &    &      & P3 & 0.0 & +20.0 & P1 $\ne$ P3? & $<$0.001 & \starxxx & P1 $>$ P3? & 1600 & \starx \\
         &    &      & P4 & 0.0 & +20.0 & P1 $\ne$ P4? & $<$0.001 & \starxxx & P1 $>$ P4? & 1600 & \starxxx \\
         & P2 & 21.3 & P3 & 0.0 & +21.3 & P2 $\ne$ P3? & $<$0.001 & \starxxx & P2 $>$ P3? & 1600 & \starxxx \\
         &    &      & P4 & 0.0 & +21.3 & P2 $\ne$ P4? & $<$0.001 & \starxxx & P2 $>$ P4? & 1600 & \starxxx \\
         & P3 & 0.0 & P4 & 0.0 & 0.0 & P3 $\ne$ P4? & 1.000 &  & - &  &  \\
        \cmidrule(r){2-12}
        \llama\ LLaMA 2
         & P1 & 11.3 & P2 & 0.0 & +11.3 & P1 $\ne$ P2? & $<$0.001 & \starxxx & P1 $>$ P2? & 1240 & \starxxx \\
         &    &      & P3 & 0.0 & +11.3 & P1 $\ne$ P3? & $<$0.001 & \starxxx & P1 $>$ P3? & 1240 & \starxxx \\
         &    &      & P4 & 3.4 & +7.9 & P1 $\ne$ P4? & $<$0.001 & \starxxx & P1 $>$ P4? & 1070 & \starxxx \\
         & P2 & 0.0 & P3 & 0.0 & 0.0 & P2 $\ne$ P3? & 1.000 &  & - &  &  \\
         &    &      & P4 & 3.4 & -3.4 & P2 $\ne$ P4? & 0.17\phantom{0} &  & - &  &  \\
         & P3 & 0.0 & P4 & 3.4 & -3.4 & P3 $\ne$ P4? & 0.17\phantom{0} &  & - &  &  \\
        \cmidrule(r){2-12}
        \mixtral\ Mixtral
         & P1 & 0.9 & P2 & 7.0 & -6.1 & P1 $\ne$ P2? & 0.054 &  & - &  &  \\
         &    &      & P3 & 27.3 & -26.4 & P1 $\ne$ P3? & $<$0.001 & \starxxx & P1 $<$ P3? & 1574 & \starxxx \\
         &    &      & P4 & 0.3 & +0.6& P1 $\ne$ P4? & 0.627 &  & - &  &  \\
         & P2 & 7.0 & P3 & 27.3 & -20.3 & P2 $\ne$ P3? & $<$0.001 & \starxxx & P2 $<$ P3? & 1520 & \starxxx \\
         &    &      & P4 & 0.3 & +6.7 & P2 $\ne$ P4? & 0.016 & \starx & P2 $>$ P4? & 1086 & \starxxx \\
         & P3 & 27.3 & P4 & 0.3 & +27.0 & P3 $\ne$ P4? & $<$0.001 & \starxxx & P3 $>$ P4? & 1579 & \starxxx \\
        \toprule
        \multicolumn{12}{r}{{\small \starx $p < 0.05$, \starxx $p < 0.01$, \starxxx $p < 0.001$}}
    \end{tabular}
    \caption{The strength of FCE, Dunn's post-test result, and Mann-Whitney U test result for H2-1, in Story 1.}
    \label{tab:h2-1-app-story1}
\end{table*}

\begin{table*}
    \centering
    \begin{tabular}{@{}lcrcrrcr@{}lcr@{}l@{}}
        \toprule
         \multicolumn{12}{l}{\textbf{Story 1. Term paper}}\\
        \toprule
         & \multicolumn{2}{l}{Condition 1} & \multicolumn{2}{l}{Condition 2} & Diff. & \multicolumn{3}{l}{Dunn's post-test} & \multicolumn{3}{l}{Mann-Whitney test} \\
         \cmidrule(r){2-3}\cmidrule(r){4-5}\cmidrule(r){7-9}\cmidrule(r){10-12}
         &  & FCE &  & FCE &  & Hypothesis & $p$-value & & Hypothesis & $U$ & \\
         \cmidrule(r){2-12}
        \gpt\ GPT-4 
         & R1 & 0.3 & R2 & 0.0 & +0.3 & R1 $\ne$ R2? & 0.848 &  & - &  &  \\
         &    &      & R3 & -20.0 & +20.3 & R1 $\ne$ R3? & $<$0.001 & \starxxx & R1 $>$ R3? & 1600 & \starxxx \\
         &    &      & R4 & 0.0 & +0.3 & R1 $\ne$ R4? & 0.848 &  & - &  &  \\
         & R2 & 0.0 & R3 & -20.0 & +20.0 & R2 $\ne$ R3? & $<$0.001 & \starxxx & R2 $>$ R3? & 1600 & \starxxx \\
         &    &      & R4 & 0.0 & +0.0 & R2 $\ne$ R4? & 1.000 &  & - &  &  \\
         & R3 & -20.0 & R4 & 0.0 & -20.0 & R3 $\ne$ R4? & $<$0.001 & \starxxx & R3 $<$ R4? & 1600 & \starxxx \\
        \cmidrule(r){2-12}
        \claude\ Claude 3
         & R1 & 20.0 & R2 & 13.5 & +6.5 & R1 $\ne$ R2? & 0.042 & \starxx & R1 $>$ R2? & 1060 & \starxxx \\
         &    &      & R3 & 1.3 & +18.7 & R1 $\ne$ R3? & $<$0.001 & \starxxx & R1 $>$ R3? & 1560 & \starxxx \\
         &    &      & R4 & -40.0 & +60.0 & R1 $\ne$ R4? & $<$0.001 & \starxxx & R1 $>$ R4? & 1600 & \starxxx \\
         & R2 & 13.5 & R3 & 1.3 & +12.2 & R2 $\ne$ R3? & $<$0.001 & \starxxx & R2 $>$ R3? & 1293 & \starxxx \\
         &    &      & R4 & -40.0 & +53.5 & R2 $\ne$ R4? & $<$0.001 & \starxxx & R2 $>$ R4? & 1600 & \starxxx \\
         & R3 & 1.3 & R4 & -40.0 & +41.3 & R3 $\ne$ R4? & $<$0.001 & \starxxx & R3 $>$ R4? & 1600 & \starxxx \\
        \cmidrule(r){2-12}
        \llama\ LLaMA 2
         & R1 & 11.3 & R2 & 7.9 & +3.4 & R1 $\ne$ R2? & 0.316 &  & - &  &  \\
         &    &      & R3 & 4.9 & +6.4 & R1 $\ne$ R3? & 0.068 &  & - &  &  \\
         &    &      & R4 & -0.7 & +12.0 & R1 $\ne$ R4? & 0.003 & \starxx & R1 $>$ R4? & 1132 & \starxxx \\
         & R2 & 7.9 & R3 & 4.9 & +3.0 & R2 $\ne$ R3? & 0.388 &  & - &  &  \\
         &    &      & R4 & -0.7 & +8.6 & R2 $\ne$ R4? & 0.05\phantom{0} &  & - &  &  \\
         & R3 & 4.9 & R4 & -0.7 & +5.6 & R3 $\ne$ R4? & 0.332 &  & - &  &  \\
        \cmidrule(r){2-12}
        \mixtral\ Mixtral
         & R1 & 0.9 & R2 & 7.3 & -6.4 & R1 $\ne$ R2? & 0.023 & \starxx & R1 $<$ R2? & 1046 & \starxx \\
         &    &      & R3 & 6.6 & -5.7 & R1 $\ne$ R3? & 0.026 & \starxx & R1 $<$ R3? & 1059 & \starxx \\
         &    &      & R4 & 5.9 & -5.0 & R1 $\ne$ R4? & 0.069 &  & - & - &  \\
         & R2 & 7.3 & R3 & 6.6 & +0.7 & R2 $\ne$ R3? & 0.955 &  & - & - &  \\
         &    &      & R4 & 5.9 & +1.4 & R2 $\ne$ R4? & 0.647 &  & - & - &  \\
         & R3 & 6.6 & R4 & 5.9 & +0.7 & R3 $\ne$ R4? & 0.688 &  & - & - &  \\
        \toprule
        \multicolumn{12}{r}{{\small \starx $p < 0.05$, \starxx $p < 0.01$, \starxxx $p < 0.001$}}
    \end{tabular}
    \caption{The strength of FCE, Dunn's post-test result, and Mann-Whitney U test result for H2-2, in Story 1.}
    \label{tab:h2-2-app-story1}
\end{table*}

%% file: table/story2.tex
\begin{table*}[p]
    \centering
    \begin{tabular}{@{}lcrcrrcr@{}lcr@{}l@{}}
        \toprule
         \multicolumn{12}{l}{\textbf{Story 2. Supermarket}}\\
        \toprule
         & \multicolumn{2}{l}{Condition 1} & \multicolumn{2}{l}{Condition 2} & Diff. & \multicolumn{3}{l}{Dunn's post-test} & \multicolumn{3}{l}{Mann-Whitney test} \\
         \cmidrule(r){2-3}\cmidrule(r){4-5}\cmidrule(r){7-9}\cmidrule(r){10-12}
         &  & FCE &  & FCE &  & Hypothesis & $p$-value & & Hypothesis & $U$ & \\
         \cmidrule(r){2-12}
        \gpt\ GPT-4 
         & P1 & 8.3 & P2 & 20.0 & -11.7 & P1 $\ne$ P2? & $<$0.001 & \starxxx & P1 $<$ P2? & 1280 & \starxxx \\
         &    &      & P3 & -9.5 & +17.8 & P1 $\ne$ P3? & $<$0.001 & \starxxx & P1 $>$ P3? & 1358 & \starxxx \\
         &    &      & P4 & 19.5 & -11.2 & P1 $\ne$ P4? & $<$0.001 & \starxxx & P1 $<$ P4? & 1259 & \starxxx \\
         & P2 & 20.0 & P3 & -9.5 & +29.5 & P2 $\ne$ P3? & $<$0.001 & \starxxx & P2 $>$ P3? & 1600 & \starxxx \\
         &    &      & P4 & 19.5 & +0.5 & P2 $\ne$ P4? & 0.845 &  & - &  &  \\
         & P3 & -9.5 & P4 & 19.5 & -29.0 & P3 $\ne$ P4? & $<$0.001 & \starxxx & P3 $<$ P4? & 1589 & \starxxx \\
        \cmidrule(r){2-12}
        \claude\ Claude 3
         & P1 & 25.5 & P2 & 29.0 & -3.5 & P1 $\ne$ P2? & 0.114 &  & - &  &  \\
         &    &      & P3 & -29.5 & +55.0 & P1 $\ne$ P3? & $<$0.001 & \starxxx & P1 $>$ P3? & 1600 & \starxxx \\
         &    &      & P4 & 27.7 & -2.2 & P1 $\ne$ P4? & 0.246 &  & - &  &  \\
         & P2 & 29.0 & P3 & -29.5 & +58.5 & P2 $\ne$ P3? & $<$0.001 & \starxxx & P2 $>$ P3? & 1600 & \starxxx \\
         &    &      & P4 & 27.7 & +1.3 & P2 $\ne$ P4? & 0.994 &  & - &  &  \\
         & P3 & -29.5 & P4 & 27.7 & -57.2 & P3 $\ne$ P4? & $<$0.001 & \starxxx & P3 $<$ P4? & 600 & \starxxx \\
        \cmidrule(r){2-12}
        \llama\ LLaMA 2
         & P1 & 1.0 & P2 & 20.0 & -19.0 & P1 $\ne$ P2? & $<$0.001 & \starxxx & P1 $<$ P2? & 1600 & \starxxx \\
         &    &      & P3 & -3.3 & +4.3 & P1 $\ne$ P3? & 0.016 & \starx & P1 $>$ P3? & 1114 & \starxxx \\
         &    &      & P4 & 37.0 & -36.0 & P1 $\ne$ P4? & 0.248 &  & - &  &  \\
         & P2 & 20.0 & P3 & -3.3 & +23.3 & P2 $\ne$ P3? & $<$0.001 & \starxxx & P2 $>$ P3? & 1600 & \starxxx \\
         &    &      & P4 & 37.0 & -17.0 & P2 $\ne$ P4? & $<$0.001 & \starxxx & P2 $<$ P4? & 1600 & \starxxx \\
         & P3 & -3.3 & P4 & 37.0 & -40.3 & P3 $\ne$ P4? & $<$0.001 & \starxxx & P3 $<$ P4? & 1222 & \starxxx \\
        \cmidrule(r){2-12}
        \mixtral\ Mixtral
         & P1 & 40.5 & P2 & 47.9 & -7.4 & P1 $\ne$ P2? & 0.271 &  & - &  &  \\
         &    &      & P3 & -33.0 & +73.5 & P1 $\ne$ P3? & $<$0.001 & \starxxx & P1 $>$ P3? & 1600 & \starxxx \\
         &    &      & P4 & 40.0 & +0.5 & P1 $\ne$ P4? & 0.437 &  & - &  &  \\
         & P2 & 47.9 & P3 & -33.0 & +80.9 & P2 $\ne$ P3? & $<$0.001 & \starxxx & P2 $>$ P3? & 1600 & \starxxx \\
         &    &      & P4 & 40.0 & +7.9 & P2 $\ne$ P4? & 0.746 &  & - &  &  \\
         & P3 & -33.0 & P4 & 40.0 & -73.3 & P3 $\ne$ P4? & $<$0.001 & \starxxx & P3 $<$ P4? & 1600 & \starxxx \\
        \bottomrule
        \multicolumn{12}{r}{{\small \starx $p < 0.05$, \starxx $p < 0.01$, \starxxx $p < 0.001$}}
    \end{tabular}
    \caption{The strength of FCE, Dunn's post-test result, and Mann-Whitney U test result for H2-1, in Story 2.}
    \label{tab:h2-1-app-story2}
\end{table*}

\begin{table*}[p]
    \centering
    \begin{tabular}{@{}lcrcrrcr@{}lcr@{}l@{}}
        \toprule
         \multicolumn{12}{l}{\textbf{Story 2. Supermarket}}\\
        \toprule
         & \multicolumn{2}{l}{Condition 1} & \multicolumn{2}{l}{Condition 2} & Diff. & \multicolumn{3}{l}{Dunn's post-test} & \multicolumn{3}{l}{Mann-Whitney test} \\
         \cmidrule(r){2-3}\cmidrule(r){4-5}\cmidrule(r){7-9}\cmidrule(r){10-12}
         &  & FCE &  & FCE &  & Hypothesis & $p$-value & & Hypothesis & $U$ & \\
         \cmidrule(r){2-12}
        \gpt\ GPT-4 
         & R1 & 8.3 & R2 & 9.5 & -1.2 & R1 $\ne$ R2? & 0.352 &  & - &  &  \\
         &    &      & R3 & 8.0 & +0.3 & R1 $\ne$ R3? & 0.998 &  & - &  &  \\
         &    &      & R4 & 24.5 & -16.2 & R1 $\ne$ R4? & $<$0.001 & \starxxx & R1 $<$ R4? & 1222 & \starxxx \\
         & R2 & 9.5 & R3 & 8.0 & +1.5 & R2 $\ne$ R3? & 0.354 &  & - &  &  \\
         &    &      & R4 & 24.5 & -15.0 & R2 $\ne$ R4? & $<$0.001 & \starxxx & R2 $<$ R4? & 1173 & \starxxx \\
         & R3 & 8.0 & R4 & 24.5 & -16.5 & R3 $\ne$ R4? & $<$0.001 & \starxxx & R3 $<$ R4? & 1223 & \starxxx \\
        \cmidrule(r){2-12}
        \claude\ Claude 3
         & R1 & 25.2 & R2 & 25.5 & -0.3 & R1 $\ne$ R2? & 0.824 &  & - &  &  \\
         &    &      & R3 & 6.6 & +18.6 & R1 $\ne$ R3? & $<$0.001 & \starxxx & R1 $>$ R3? & 1489 & \starxxx \\
         &    &      & R4 & 2.3 & +22.9 & R1 $\ne$ R4? & $<$0.001 & \starxxx & R1 $>$ R4? & 1600 & \starxxx \\
         & R2 & 25.5 & R3 & 6.6 & +18.9 & R2 $\ne$ R3? & $<$0.001 & \starxxx & R2 $>$ R3? & 1440 & \starxxx \\
         &    &      & R4 & 2.3 & +23.2 & R2 $\ne$ R4? & $<$0.001 & \starxxx & R2 $>$ R4? & 1526 & \starxxx \\
         & R3 & 6.6 & R4 & 2.3 & +4.3 & R3 $\ne$ R4? & 0.854 &  & - &  &  \\
        \cmidrule(r){2-12}
        \llama\ LLaMA 2
         & R1 & 1.0 & R2 & -0.9 & +1.9 & R1 $\ne$ R2? & 0.678 &  & - &  &  \\
         &    &      & R3 & 2.8 & -1.8 & R1 $\ne$ R3? & 0.432 &  & - &  &  \\
         &    &      & R4 & 6.5 & -5.5 & R1 $\ne$ R4? & 0.194 &  & - &  &  \\
         & R2 & -0.9 & R3 & 2.8 & -3.7 & R2 $\ne$ R3? & 0.237 &  & - &  &  \\
         &    &      & R4 & 6.5 & -7.4 & R2 $\ne$ R4? & 0.092 &  & - &  &  \\
         & R3 & 2.8 & R4 & 6.5 & -3.7 & R3 $\ne$ R4? & 0.611 &  & - &  &  \\
        \cmidrule(r){2-12}
        \mixtral\ Mixtral
         & R1 & 40.5 & R2 & 32.6 & +7.9 & R1 $\ne$ R2? & 0.012 & \starx & R1 $>$ R2? & 1144 & \starxxx \\
         &    &      & R3 & 14.6 & +25.9 & R1 $\ne$ R3? & $<$0.001 & \starxxx & R1 $>$ R3? & 1387 & \starxxx \\
         &    &      & R4 & 46.6 & -6.1 & R1 $\ne$ R4? & 0.117 &  & - &  &  \\
         & R2 & 32.6 & R3 & 14.6 & +18.0 & R2 $\ne$ R3? & $<$0.001 & \starxxx & R2 $>$ R3? & 1375 & \starxxx \\
         &    &      & R4 & 46.6 & -14.0 & R2 $\ne$ R4? & $<$0.001 & \starxxx & R2 $<$ R4? & 1306 & \starxxx \\
         & R3 & 14.6 & R4 & 46.6 & -32.0 & R3 $\ne$ R4? & $<$0.001 & \starxxx & R3 $<$ R4? & 1528 & \starxxx \\
        \bottomrule
        \multicolumn{12}{r}{{\small \starx $p < 0.05$, \starxx $p < 0.01$, \starxxx $p < 0.001$}}
    \end{tabular}
    \caption{The strength of FCE, Dunn's post-test result, and Mann-Whitney U test result for H2-2, in Story 2.}
    \label{tab:h2-2-app-story2}
\end{table*}

%% file: table/story3.tex
\begin{table*}[p]
    \centering
    \begin{tabular}{@{}lcrcrrcr@{}lcr@{}l@{}}
        \toprule
         \multicolumn{12}{l}{\textbf{Story 3. Traffic Ticket}}\\
        \toprule
         & \multicolumn{2}{l}{Condition 1} & \multicolumn{2}{l}{Condition 2} & Diff. & \multicolumn{3}{l}{Dunn's post-test} & \multicolumn{3}{l}{Mann-Whitney test} \\
         \cmidrule(r){2-3}\cmidrule(r){4-5}\cmidrule(r){7-9}\cmidrule(r){10-12}
         &  & FCE &  & FCE &  & Hypothesis & $p$-value & & Hypothesis & $U$ & \\
         \cmidrule(r){2-12}
        \gpt\ GPT-4 
         & P1 & 2.5 & P2 & 10.0 & -7.5 & P1 $\ne$ P2? & $<$0.001 & \starxxx & P1 $<$ P2? & 1400 & \starxxx \\
         &    &      & P3 & -9.5 & +12.0 & P1 $\ne$ P3? & $<$0.001 & \starxxx & P1 $>$ P3? & 1570 & \starxxx \\
         &    &      & P4 & 10.0 & -7.5 & P1 $\ne$ P4? & $<$0.001 & \starxxx & P1 $<$ P4? & 1400 & \starxxx \\
         & P2 & 10.0 & P3 & -9.5 & +19.5 & P2 $\ne$ P3? & $<$0.001 & \starxxx & P2 $>$ P3? & 1600 & \starxxx \\
         &    &      & P4 & 10.0 & +0.0 & P2 $\ne$ P4? & 1.000 &  & P2 $<$ P4? & - &  \\
         & P3 & -9.5 & P4 & 10.0 & -19.5 & P3 $\ne$ P4? & $<$0.001 & \starxxx & P3 $<$ P4? & 1600 & \starxxx \\
        \cmidrule(r){2-12}
        \claude\ Claude 3
         & P1 & 0.0 & P2 & 0.0 & +0.0 & P1 $\ne$ P2? & 1.000 &  & P1 $<$ P2? & - &  \\
         &    &      & P3 & 0.0 & +0.0 & P1 $\ne$ P3? & 1.000 &  & P1 $<$ P3? & - &  \\
         &    &      & P4 & 0.0 & +0.0 & P1 $\ne$ P4? & 1.000 &  & P1 $<$ P4? & - &  \\
         & P2 & 0.0 & P3 & 0.0 & +0.0 & P2 $\ne$ P3? & 1.000 &  & P2 $<$ P3? & - &  \\
         &    &      & P4 & 0.0 & +0.0 & P2 $\ne$ P4? & 1.000 &  & P2 $<$ P4? & - &  \\
         & P3 & 0.0 & P4 & 0.0 & +0.0 & P3 $\ne$ P4? & 1.000 &  & P3 $<$ P4? & - &  \\
        \cmidrule(r){2-12}
        \llama\ LLaMA 2
         & P1 & 0.0 & P2 & 10.0 & -10.0 & P1 $\ne$ P2? & $<$0.001 & \starxxx & P1 $<$ P2? & 1600 & \starxxx \\
         &    &      & P3 & 0.0 & +0.0 & P1 $\ne$ P3? & 1.000 & \starxxx & P1 $<$ P3? & - &  \\
         &    &      & P4 & 0.0 & +0.0 & P1 $\ne$ P4? & 1.000 & \starxxx & P1 $<$ P4? & - &  \\
         & P2 & 10.0 & P3 & 0.0 & +10.0 & P2 $\ne$ P3? & $<$0.001 & \starxxx & P2 $>$ P3? & 1600 & \starxxx \\
         &    &      & P4 & 0.0 & +10.0 & P2 $\ne$ P4? & $<$0.001 & \starxxx & P2 $>$ P4? & 1600 & \starxxx \\
         & P3 & 0.0 & P4 & 0.0 & +0.0 & P3 $\ne$ P4? & 1.000 &  & P3 $<$ P4? & - &  \\
        \cmidrule(r){2-12}
        \mixtral\ Mixtral
         & P1 & 9.3 & P2 & 10.5 & -1.2 & P1 $\ne$ P2? & 0.694 &  & P1 $<$ P2? & - &  \\
         &    &      & P3 & -31.0 & +40.3 & P1 $\ne$ P3? & $<$0.001 & \starxxx & P1 $>$ P3? & 1583 & \starxxx \\
         &    &      & P4 & 9.0 & +0.3 & P1 $\ne$ P4? & 0.833 &  & P1 $<$ P4? & - &  \\
         & P2 & 10.5 & P3 & -31.0 & +41.5 & P2 $\ne$ P3? & $<$0.001 & \starxxx & P2 $>$ P3? & 1600 & \starxxx \\
         &    &      & P4 & 9.0 & +1.5 & P2 $\ne$ P4? & 0.545 &  & P2 $<$ P4? & - &  \\
         & P3 & -31.0 & P4 & 9.0 & -40.0 & P3 $\ne$ P4? & $<$0.001 & \starxxx & P3 $<$ P4? & 1590 & \starxxx \\
        \bottomrule
        \multicolumn{12}{r}{{\small \starx $p < 0.05$, \starxx $p < 0.01$, \starxxx $p < 0.001$}}
    \end{tabular}
    \caption{The strength of FCE, Dunn's post-test result, and Mann-Whitney U test result for H2-1, in Story 3.}
    \label{tab:h2-1-app-story3}
\end{table*}

\begin{table*}[p]
    \centering
    \begin{tabular}{@{}lcrcrrcr@{}lcr@{}l@{}}
        \toprule
         \multicolumn{12}{l}{\textbf{Story 3. Traffic Ticket}}\\
        \toprule
         & \multicolumn{2}{l}{Condition 1} & \multicolumn{2}{l}{Condition 2} & Diff. & \multicolumn{3}{l}{Dunn's post-test} & \multicolumn{3}{l}{Mann-Whitney test} \\
         \cmidrule(r){2-3}\cmidrule(r){4-5}\cmidrule(r){7-9}\cmidrule(r){10-12}
         &  & FCE &  & FCE &  & Hypothesis & $p$-value & & Hypothesis & $U$ & \\
         \cmidrule(r){2-12}
        \gpt\ GPT-4 
         & R1 & 2.5 & R2 & 10.0 & -7.5 & R1 $\ne$ R2? & $<$0.001 & \starxxx & R1 $<$ R2? & 1400 & \starxxx \\
         &    &      & R3 & 0.0 & +2.5 & R1 $\ne$ R3? & 0.854 &  & - &  &  \\
         &    &      & R4 & -2.8 & +5.3 & R1 $\ne$ R4? & 0.001 & \starxx & R1 $>$ R4? & 1175 & \starxxx \\
         & R2 & 10.0 & R3 & 0.0 & +10.0 & R2 $\ne$ R3? & $<$0.001 & \starxxx & R2 $>$ R3? & 1600 & \starxxx \\
         &    &      & R4 & -2.8 & +12.8 & R2 $\ne$ R4? & $<$0.001 & \starxxx & R2 $>$ R4? & 1340 & \starxxx \\
         & R3 & 0.0 & R4 & -2.8 & +2.8 & R3 $\ne$ R4? & 0.145 &  & - &  &  \\
        \cmidrule(r){2-12}
        \claude\ Claude 3
         & R1 & 0.0 & R2 & 0.0 & +0.0 & R1 $\ne$ R2? & 1.000 &  & - &  &  \\
         &    &      & R3 & 4.9 & -4.9 & R1 $\ne$ R3? & 0.001 & \starxx & R1 $<$ R3? & 1180 & \starxxx \\
         &    &      & R4 & -48.8 & +48.8 & R1 $\ne$ R4? & $<$0.001 & \starxxx & R1 $>$ R4? & 1600 & \starxxx \\
         & R2 & 0.0 & R3 & 4.9 & -4.9 & R2 $\ne$ R3? & 0.001 & \starxx & R2 $<$ R3? & 1180 & \starxxx \\
         &    &      & R4 & -48.8 & +48.8 & R2 $\ne$ R4? & $<$0.001 & \starxxx & R2 $>$ R4? & 1600 & \starxxx \\
         & R3 & 4.9 & R4 & -48.8 & +53.7 & R3 $\ne$ R4? & $<$0.001 & \starxxx & R3 $>$ R4? & 1600 & \starxxx \\
        \cmidrule(r){2-12}
        \llama\ LLaMA 2
         & R1 & 0.0 & R2 & 0.1 & -0.1 & R1 $\ne$ R2? & 0.899 &  & - &  &  \\
         &    &      & R3 & -8.4 & +8.4 & R1 $\ne$ R3? & 0.252 &  & - &  &  \\
         &    &      & R4 & 21.1 & -21.1 & R1 $\ne$ R4? & $<$0.001 & \starxxx & R1 $<$ R4? & 1240 & \starxxx \\
         & R2 & 0.1 & R3 & -8.4 & +8.5 & R2 $\ne$ R3? & 0.22\phantom{0} &  & - &  &  \\
         &    &      & R4 & 21.1 & -21.0 & R2 $\ne$ R4? & $<$0.001 & \starxxx & R2 $<$ R4? & 1082 & \starxxx \\
         & R3 & -8.4 & R4 & 21.1 & -29.5 & R3 $\ne$ R4? & $<$0.001 & \starxxx & R3 $<$ R4? & 647 & \starxxx \\
        \cmidrule(r){2-12}
        \mixtral\ Mixtral
         & R1 & 9.3 & R2 & 6.0 & +3.3 & R1 $\ne$ R2? & 0.127 &  & - &  &  \\
         &    &      & R3 & 5.9 & +3.4 & R1 $\ne$ R3? & 0.18\phantom{0} &  & - &  &  \\
         &    &      & R4 & 19.9 & -10.6 & R1 $\ne$ R4? & $<$0.001 & \starxxx & R1 $<$ R4? & 1225 & \starxxx \\
         & R2 & 6.0 & R3 & 5.9 & +0.1 & R2 $\ne$ R3? & 0.854 &  & - &  &  \\
         &    &      & R4 & 19.9 & -13.9 & R2 $\ne$ R4? & $<$0.001 & \starxxx & R2 $<$ R4? & 1425 & \starxxx \\
         & R3 & 5.9 & R4 & 19.9 & -14.0 & R3 $\ne$ R4? & $<$0.001 & \starxxx & R3 $<$ R4? & 1316 & \starxxx \\
        \bottomrule
        \multicolumn{12}{r}{{\small \starx $p < 0.05$, \starxx $p < 0.01$, \starxxx $p < 0.001$}}
    \end{tabular}
    \caption{The strength of FCE, Dunn's post-test result, and Mann-Whitney U test result for H2-2, in Story 3.}
    \label{tab:h2-2-app-story3}
\end{table*}

%% file: table/story4.tex
\begin{table*}[p]
    \centering
    \begin{tabular}{@{}lcrcrrcr@{}lcr@{}l@{}}
        \toprule
         \multicolumn{12}{l}{\textbf{Story 4. Space R\&D Program}}\\
        \toprule
         & \multicolumn{2}{l}{Condition 1} & \multicolumn{2}{l}{Condition 2} & Diff. & \multicolumn{3}{l}{Dunn's post-test} & \multicolumn{3}{l}{Mann-Whitney test} \\
         \cmidrule(r){2-3}\cmidrule(r){4-5}\cmidrule(r){7-9}\cmidrule(r){10-12}
         &  & FCE &  & FCE &  & Hypothesis & $p$-value & & Hypothesis & $U$ & \\
         \cmidrule(r){2-12}
        \gpt\ GPT-4 
         & P1 & 20.0 & P2 & 20.0 & +0.0 & P1 $\ne$ P2? & 1.000 &  & - &  &  \\
         &    &      & P3 & 0.0 & +20.0 & P1 $\ne$ P3? & $<$0.001 & \starxxx & P1 $>$ P3? & 1600 & \starxxx \\
         &    &      & P4 & 20.0 & +0.0 & P1 $\ne$ P4? & 1.000 &  & - &  &  \\
         & P2 & 20.0 & P3 & 0.0 & +20.0 & P2 $\ne$ P3? & $<$0.001 & \starxxx & P2 $>$ P3? & 1600 & \starxxx \\
         &    &      & P4 & 20.0 & +0.0 & P2 $\ne$ P4? & 1.000 &  & - &  &  \\
         & P3 & 0.0 & P4 & 20.0 & -20.0 & P3 $\ne$ P4? & $<$0.001 & \starxxx & P3 $<$ P4? & 1600 & \starxxx \\
        \cmidrule(r){2-12}
        \claude\ Claude 3
         & P1 & 9.1 & P2 & 5.0 & +4.1 & P1 $\ne$ P2? & 0.114 &  & - &  &  \\
         &    &      & P3 & 0.0 & +9.1 & P1 $\ne$ P3? & $<$0.001 & \starxxx & P1 $>$ P3? & 1160 & \starxxx \\
         &    &      & P4 & 20.4 & -11.3 & P1 $\ne$ P4? & $<$0.001 & \starxxx & P1 $<$ P4? & 650 & \starxxx \\
         & P2 & 5.0 & P3 & 0.0 & +5.0 & P2 $\ne$ P3? & 0.058 &  & - &  &  \\
         &    &      & P4 & 20.4 & -15.4 & P2 $\ne$ P4? & $<$0.001 & \starxxx & P2 $<$ P4? & 740 & \starxxx \\
         & P3 & 0.0 & P4 & 20.4 & -20.4 & P3 $\ne$ P4? & $<$0.001 & \starxxx & P3 $<$ P4? & 840 & \starxxx \\
        \cmidrule(r){2-12}
        \llama\ LLaMA 2
         & P1 & 17.8 & P2 & 20.0 & -2.2 & P1 $\ne$ P2? & 0.489 &  & - &  &  \\
         &    &      & P3 & -19.0 & +36.8 & P1 $\ne$ P3? & $<$0.001 & \starxxx & P1 $>$ P3? & 1595 & \starxxx \\
         &    &      & P4 & 12.5 & +5.3 & P1 $\ne$ P4? & 0.073 &  & - &  &  \\
         & P2 & 20.0 & P3 & -19.0 & +39.0 & P2 $\ne$ P3? & $<$0.001 & \starxxx & P2 $>$ P3? & 1600 & \starxxx \\
         &    &      & P4 & 12.5 & +7.5 & P2 $\ne$ P4? & 0.013 & \starx & P2 $>$ P4? & 1100 & \starxxx \\
         & P3 & -19.0 & P4 & 12.5 & -31.5 & P3 $\ne$ P4? & $<$0.001 & \starxxx & P3 $<$ P4? & 1585 & \starxxx \\
        \cmidrule(r){2-12}
        \mixtral\ Mixtral
         & P1 & 3.3 & P2 & 0.0 & +3.3 & P1 $\ne$ P2? & 0.014 & \starx & P1 $>$ P2? & 1120 & \starxxx \\
         &    &      & P3 & 1.6 & +1.7 & P1 $\ne$ P3? & 0.016 & \starx & P1 $>$ P3? & 958 & \starx \\
         &    &      & P4 & 0.5 & +2.8 & P1 $\ne$ P4? & $<$0.001 & \starxxx & P1 $>$ P4? & 1022 & \starx \\
         & P2 & 0.0 & P3 & 1.6 & -1.6 & P2 $\ne$ P3? & 0.289 &  & - &  &  \\
         &    &      & P4 & 0.5 & -0.5 & P2 $\ne$ P4? & 0.243 &  & - &  &  \\
         & P3 & 1.6 & P4 & 0.5 & +1.1 & P3 $\ne$ P4? & 0.026 & \starxx & P3 $>$ P4? & 1017 & \starx \\
        \bottomrule
        \multicolumn{12}{r}{{\small \starx $p < 0.05$, \starxx $p < 0.01$, \starxxx $p < 0.001$}}
    \end{tabular}
    \caption{The strength of FCE, Dunn's post-test result, and Mann-Whitney U test result for H2-1, in Story 4.}
    \label{tab:h2-1-app-story4}
\end{table*}

\begin{table*}[p]
    \centering
    \begin{tabular}{@{}lcrcrrcr@{}lcr@{}l@{}}
        \toprule
         \multicolumn{12}{l}{\textbf{Story 4. Space R\&D Program}}\\
        \toprule
         & \multicolumn{2}{l}{Condition 1} & \multicolumn{2}{l}{Condition 2} & Diff. & \multicolumn{3}{l}{Dunn's post-test} & \multicolumn{3}{l}{Mann-Whitney test} \\
         \cmidrule(r){2-3}\cmidrule(r){4-5}\cmidrule(r){7-9}\cmidrule(r){10-12}
         &  & FCE &  & FCE &  & Hypothesis & $p$-value & & Hypothesis & $U$ & \\
         \cmidrule(r){2-12}
        \gpt\ GPT-4 
         & R1 & 20.0 & R2 & 20.0 & +0.0 & R1 $\ne$ R2? & 1.000 &  & - &  &  \\
         &    &      & R3 & 0.3 & +19.7 & R1 $\ne$ R3? & 0.827 &  &  &  &  \\
         &    &      & R4 & 7.5 & +12.5 & R1 $\ne$ R4? & $<$0.001 & \starxxx & R1 $>$ R4? & 1400 & \starxxx \\
         & R2 & 20.0 & R3 & 0.3 & +19.7 & R2 $\ne$ R3? & 0.827 &  & - &  &  \\
         &    &      & R4 & 7.5 & +12.5 & R2 $\ne$ R4? & $<$0.001 & \starxxx & R2 $>$ R4? & 1400 & \starxxx \\
         & R3 & 0.3 & R4 & 7.5 & -7.2 & R3 $\ne$ R4? & $<$0.001 & \starxxx & R3 $<$ R4? & 1404 & \starxxx \\
        \cmidrule(r){2-12}
        \claude\ Claude 3
         & R1 & 9.1 & R2 & 20.3 & -11.2 & R1 $\ne$ R2? & $<$0.001 & \starxxx & R1 $<$ R2? & 1229 & \starxxx \\
         &    &      & R3 & 15.9 & -6.8 & R1 $\ne$ R3? & 0.04\phantom{0} & \starx & R1 $<$ R3? & 1013 & \starxxx \\
         &    &      & R4 & -0.9 & +10.0 & R1 $\ne$ R4? & $<$0.001 & \starxxx & R1 $>$ R4? & 1182 & \starxxx \\
         & R2 & 20.3 & R3 & 15.9 & +4.4 & R2 $\ne$ R3? & 0.002 & \starxx & R2 $>$ R3? & 1268 & \starxxx \\
         &    &      & R4 & -0.9 & +21.2 & R2 $\ne$ R4? & $<$0.001 & \starxxx & R2 $>$ R4? & 1600 & \starxxx \\
         & R3 & 15.9 & R4 & -0.9 & +16.8 & R3 $\ne$ R4? & $<$0.001 & \starxxx & R3 $>$ R4? & 1579 & \starxxx \\
        \cmidrule(r){2-12}
        \llama\ LLaMA 2
         & R1 & 17.8 & R2 & 11.5 & +6.3 & R1 $\ne$ R2? & $<$0.001 & \starxxx & R1 $>$ R2? & 1362 & \starxxx \\
         &    &      & R3 & 2.1 & +15.7 & R1 $\ne$ R3? & $<$0.001 & \starxxx & R1 $>$ R3? & 1431 & \starxxx \\
         &    &      & R4 & 5.8 & +12.0 & R1 $\ne$ R4? & $<$0.001 & \starxxx & R1 $>$ R4? & 976 & \starxx \\
         & R2 & 11.5 & R3 & 2.1 & +9.4 & R2 $\ne$ R3? & 0.118 &  & - &  &  \\
         &    &      & R4 & 5.8 & +5.7 & R2 $\ne$ R4? & 0.01\phantom{0} & \starx & R2 $>$ R4? & 811 & \starx \\
         & R3 & 2.1 & R4 & 5.8 & -3.7 & R3 $\ne$ R4? & 0.288 &  & - &  &  \\
        \cmidrule(r){2-12}
        \mixtral\ Mixtral
         & R1 & 3.3 & R2 & 5.1 & -1.8 & R1 $\ne$ R2? & 0.543 &  & - &  &  \\
         &    &      & R3 & 2.7 & +0.6 & R1 $\ne$ R3? & 0.662 &  & - &  &  \\
         &    &      & R4 & 5.4 & -1.9 & R1 $\ne$ R4? & 0.667 &  & - &  &  \\
         & R2 & 5.1 & R3 & 2.7 & +2.4 & R2 $\ne$ R3? & 0.295 &  & - &  &  \\
         &    &      & R4 & 5.4 & -0.3 & R2 $\ne$ R4? & 0.858 &  & - &  &  \\
         & R3 & 2.7 & R4 & 5.4 & -2.7 & R3 $\ne$ R4? & 0.386 &  & - &  &  \\
        \bottomrule
        \multicolumn{12}{r}{{\small \starx $p < 0.05$, \starxx $p < 0.01$, \starxxx $p < 0.001$}}
    \end{tabular}
    \caption{The strength of FCE, Dunn's post-test result, and Mann-Whitney U test result for H2-2, in Story 4.}
    \label{tab:h2-2-app-story4}
\end{table*}